\documentclass[review]{elsarticle}

\usepackage[pagewise]{lineno}
\usepackage{graphicx}
\usepackage{hyperref} 
\usepackage{amssymb}
\usepackage{amsmath}
\usepackage{graphicx}
\usepackage{tikz}
\usepackage{pstricks}
\usepackage{xstring}
\usepackage{multirow}
\usepackage{hhline}
\usepackage{caption}
\usepackage{subcaption}
\usepackage{marvosym}
\usepackage{fontawesome}
\usepackage{wasysym}
\usepackage{tikzsymbols}
\usepackage{booktabs}
\usepackage{bbm}
\usepackage[utf8]{inputenc}
\usepackage{nicefrac}
\usepackage{relsize}
\usepackage{subcaption}
\usepackage{verbatim}
\usepackage[bottom]{footmisc}

\definecolor{darkblue}{rgb}{0,0,0.5}

\hypersetup{colorlinks=true,linkcolor=black,citecolor=darkblue,urlcolor=darkblue} 

\renewcommand{\bar}[1]{\,\overline{\!{#1}}} 

\newcommand{\demand}{M}                     
\newcommand{\cycle}{\tau}                   
\newcommand{\minfleet}{N_{min}}             
\newcommand{\fleetmult}{\eta}               
\newcommand{\thresh}{\gamma}                

\newcommand{\busspeed}{V_{bus}}             
\newcommand{\walkspeed}{V_{walk}}           

\newcommand{\boardpp}{\beta}                
\newcommand{\alightpp}{\alpha}              

\newcommand{\avgdist}{\bar{D}}              
\newcommand{\dist}[1]{D_{#1}}               

\newcommand{\avgcruise}{\bar{C}}            
\newcommand{\excruise}[1]{C_{{#1}}}         

\newcommand{\avgrate}{\bar{\lambda}}        
\newcommand{\rate}[1]{\lambda_{#1}}         

\newcommand{\avgprob}{\bar{p}}              
\newcommand{\prob}[1]{p_{#1}}               

\newcommand{\serve}[2]{x_{#1,#2}}           
\renewcommand{\split}[2]{y_{#1,#2}}         

\newcommand{\head}[2]{h_{#1,#2}}            
\newcommand{\dephead}[2]{\hat{h}_{#1,#2}}   
\newcommand{\cruise}[2]{c_{{#1,#2}}}        
\newcommand{\dwell}[2]{w_{#1,#2}}           
\newcommand{\arr}[2]{t_{#1,#2}}             
\newcommand{\dep}[2]{d_{#1,#2}}             
\newcommand{\board}[2]{b_{#1,#2}}           
\newcommand{\alight}[2]{a_{#1,#2}}          
\newcommand{\wboard}[2]{f_{#1,#2}}          
\newcommand{\walight}[2]{g_{#1,#2}}         
\newcommand{\load}[2]{l_{#1,#2}}            
\newcommand{\noise}[2]{\epsilon_{#1,#2}}    

\newcommand{\cruiseshape}{\kappa}           
\newcommand{\cruisescale}{\theta}           

\newcommand{\waitmult}{w_{wait}}            
\newcommand{\walkmult}{w_{walk}}            
\newcommand{\waittime}{T_{wait}}            
\newcommand{\vehtime}{T_{veh}}              
\newcommand{\walktime}{T_{walk}}            
\newcommand{\avgcost}{Q}                    
\newcommand{\expcost}{\hat{Q}}              
\newcommand{\overhead}{\phi}                    

\journal{Transportation Research Part C}

\makeatletter
\def\ps@pprintTitle{%
  \let\@oddhead\@empty
  \let\@evenhead\@empty
  \let\@oddfoot\@empty
  \let\@evenfoot\@oddfoot
}
\makeatother

\begin{document}

\begin{frontmatter}

\title{Application of Modular Vehicle Technology to Mitigate Bus Bunching}


\author[nyuad,correspondingauthor]{Zaid Saeed Khan}
\author{Weili He}
\author[nyuad]{M\'onica~Men\'endez}

\cortext[correspondingauthor]{Corresponding author, \href{mailto:zaid.khan@nyu.edu}{zaid.khan@nyu.edu}}
\address[nyuad]{Division of Engineering, New York University Abu Dhabi, Saadiyat Marina District PO Box 129188 - Abu Dhabi, United Arab Emirates}


\begin{abstract}
The stochastic nature of public transport systems leads to headway variability and bus bunching, causing both operator and passenger cost to increase significantly. Traditional strategies to counter bus bunching, including bus-holding, stop-skipping, and bus substitution/insertion, suffer from trade-offs and shortcomings. Autonomous modular vehicle (AMV) technology provides an additional level of flexibility in bus dispatching and operations, which can offer significant benefits in mitigating bus bunching compared to strategies available with conventional buses. This paper introduces a novel alternative to stop-skipping by leveraging the new capabilities offered by AMVs (in particular, en-route coupling and decoupling of modular units). We develop a simple \textit{bus-splitting} strategy that directs a modular bus to decouple into individual units when it experiences a headway longer than a given threshold. We then use a macroscopic simulation to present a proof-of-concept evaluation of the proposed modular strategy compared to a benchmark traditional stop-skipping strategy and the base (no control) case. We find that the proposed strategy outperforms the benchmark in decreasing each of the three travel time components: waiting time, in-vehicle time, and walking time (which it eliminates completely). It therefore reduces the overhead of bus bunching and thus the travel cost by more than twice as much as the benchmark for busy bus lines. Simultaneously, it also reduces headway variability to a comparable degree. Furthermore, we analyze different control thresholds for applying the proposed strategy, and show that it is most effective when applied proactively, i.e. with the control action being triggered even by small headway deviations. \vspace{12pt} 
\end{abstract}

\begin{keyword}
Bus bunching; Modular bus units; Stop skipping; Autonomous modular vehicles; Bus splitting
\end{keyword}

\end{frontmatter}

\pagebreak 


\section{Problem Description \& Literature Review}

Bus operations on typical urban networks are highly stochastic due to their interactions with car traffic, passenger demand, and signal timings, among others, which are all random processes \cite{loder2017empirics}. 
The outcome of these interactions is potentially large deviations in bus headways, causing buses to bunch together \cite{daganzo2009}. This is because the number of passengers boarding the bus is positively correlated with the bus headway \cite{newell1964}: a bus arriving earlier (resp. later) than scheduled encounters less (more) passengers on average relative to the bus before it. This, in turn, results in a shorter (longer) dwell time compared to the average, further increasing the uneven distribution of passengers across buses and the resulting headway deviation. This phenomenon is commonly referred to as bus bunching. Since travel time reliability is one of the most important criteria for selecting buses as the mode of transport, this may also result in long-term negative changes to the demand, leading to the reduced productivity of bus services and increased cost per passenger \cite{nesheli2015}.

Most scientific literature on the bus bunching problem focuses on strategies that hold buses at a subset of bus stops (known as control points) along a bus line \cite{daganzo2011, delgado2012, berrebi2018}. The objective is to increase headways that have become too short, breaking the positive feedback loop and preventing the bus from catching up and bunching with the previous (late) bus. However, a major drawback of these bus-holding strategies when implemented alone is that they cannot speed up late buses. Strategies proposed to overcome this problem include various flavors of \textit{stop-skipping} \cite{liu2013bus, niu2011determination, sun2005real} and bus substitution/insertion \cite{petit2018dynamic, morales2019stochastic}. Both these types of strategies have proven effective in mitigating bus bunching, but the former impose additional waiting and walking time on certain passengers, while the latter require a larger vehicle fleet as some buses always need to be in standby. As a result, these strategies are seldom adopted in practice \cite{menendez2021chapter}.

\begin{figure}[!t]
\centering
\includegraphics[width=\textwidth]{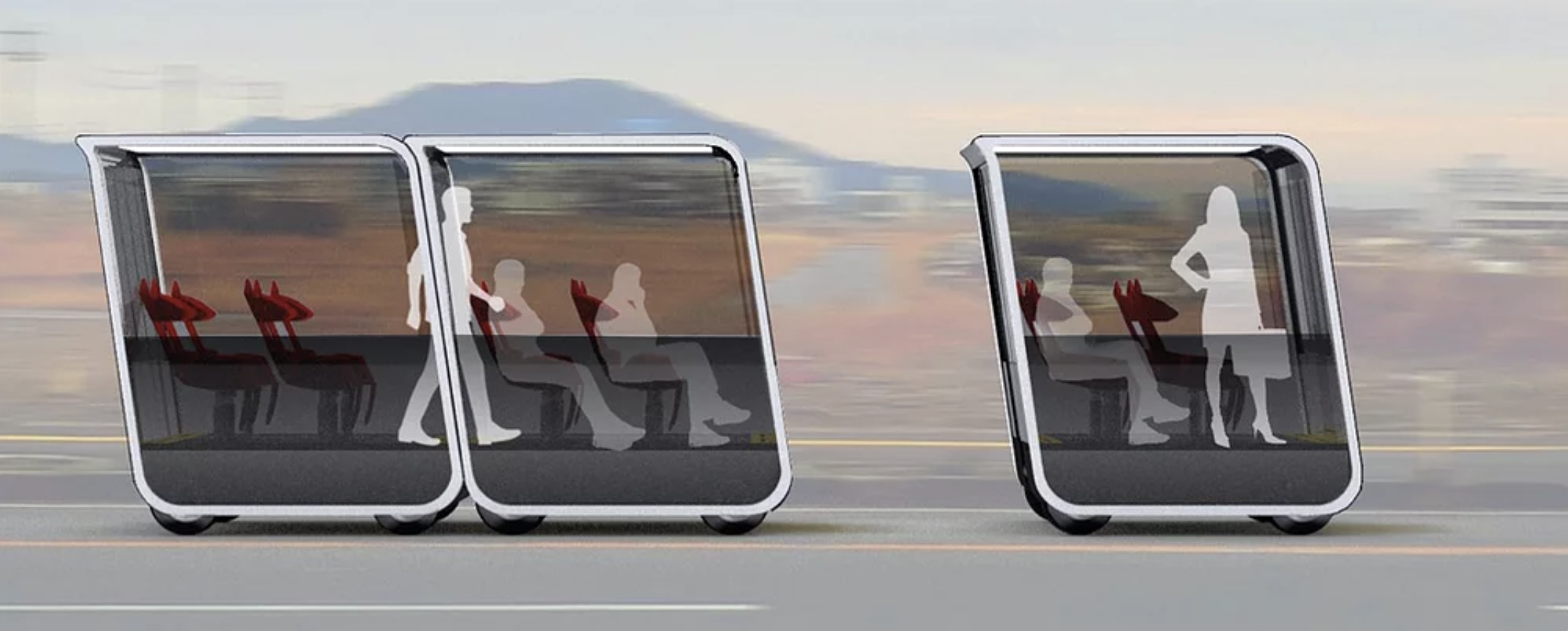}
\caption{Autonomous Modular Vehicles concept as presented by NEXT Future Transportation \cite{NextFTwebpage}}
\label{fig:Next} 
\end{figure}

In recent times, \textit{autonomous modular vehicles} (AMVs), an exciting novel technology with great potential for solving this issue, are rapidly approaching deployment. An AMV consists of modular units that can combine and split as required, with each unit or combination of units capable of operating independently. In particular, NEXT Future Transportation \cite{NextFTwebpage} have produced AMVs capable of \textit{in-motion transfer}, which allows the modular units to couple and decouple while moving on roads, so that passengers can transfer from one unit to another while traveling (Figure~\ref{fig:Next}). This provides a tremendous amount of additional flexibility that has applications in a wide range of transportation modes, including public transport, on-demand transit, ride-sharing, and emergency vehicle services, among others  \cite{gecchelin2019modular}. For bus systems, \textit{autonomous modular buses} (AMBs) have the potential to improve service frequency setting and vehicle allocation, reducing both the operating and passenger costs while improving system adaptability and reliability. Researchers have begun to explore the benefits that AMBs can bring to bus systems, and some studies already exist \citep{chen2019operational,chen2020operational,dai2020joint,dakic2021design,chen2021designing,shi2020variable,shi2021operations}. However, relatively few of the existing studies leverage the promising additional capability of in-motion transfers. Wu et al. \citep{wu2021modular} evaluated the performance benefits of an AMB system in a large-scale urban network. Caros and Chow \cite{caros2021day} used a day-to-day learning framework to compare different bus route operating strategies and quantify the benefit of in-motion transfer. Gong et al. \cite{gong2021transfer} developed a heuristic to solve a passenger-route assignment problem with a modular fleet capable of in-motion transfers.

An AMB-based bus system with in-motion transfer can potentially be very useful in solving the bus bunching problem. This paper introduces a new alternative to the traditional bus stop-skipping strategy that leverages this capability. This represents the first proof-of-concept attempt to evaluate the benefits AMBs can provide compared to the strategies available with traditional buses. To the best of our knowledge, application of AMBs to increase service reliability on fixed bus lines has not been studied before. We propose a simple \textit{bus-splitting} control policy with AMBs which is a natural extension of the traditional bus stop-skipping policy proposed by Vuchic \cite{vukan1973}. We compare our proposed policy with both the aforementioned traditional bus stop-skipping policy and a system without any control.

The remainder of this paper is structured as follows. In Section~\ref{sec:concept}, we provide a motivating example where we compare the concept of modular bus-splitting with traditional stop-skipping. In Section~\ref{sec:model}, we develop our model in detail, including the basic setting without control, the benchmark stop-skipping policy, and our bus-splitting policy. In Section~\ref{sec:experiments}, we describe our experimental settings. We present our results in Section~\ref{sec:results}, including a detailed evaluation of a representative sample as well as analysis of the policy's robustness. Finally, we offer concluding thoughts and directions for future work in Section~\ref{sec:conclusion}.


\section{Motivating Example}\label{sec:concept}

\begin{figure}[!b]\centering
        \begin{subfigure}[b]{0.48\textwidth}\centering
        \includegraphics[width=\textwidth]{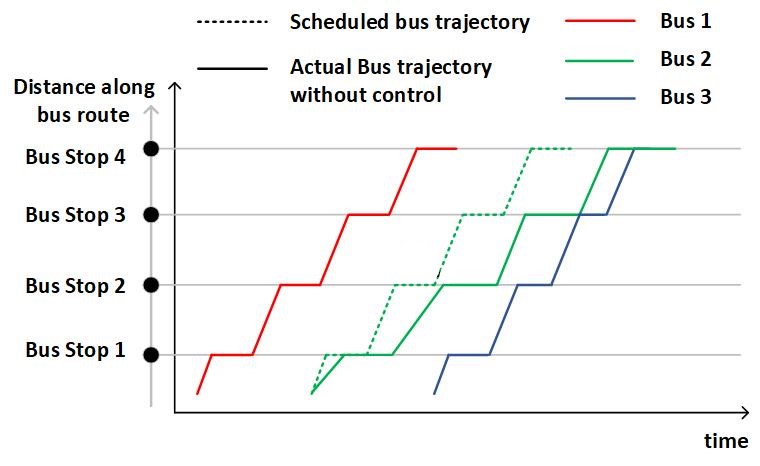}
        \caption{No control}\label{fig:nocontrolconcept}
        \end{subfigure}
        \begin{subfigure}[b]{0.48\textwidth}\centering
        \includegraphics[width=\textwidth]{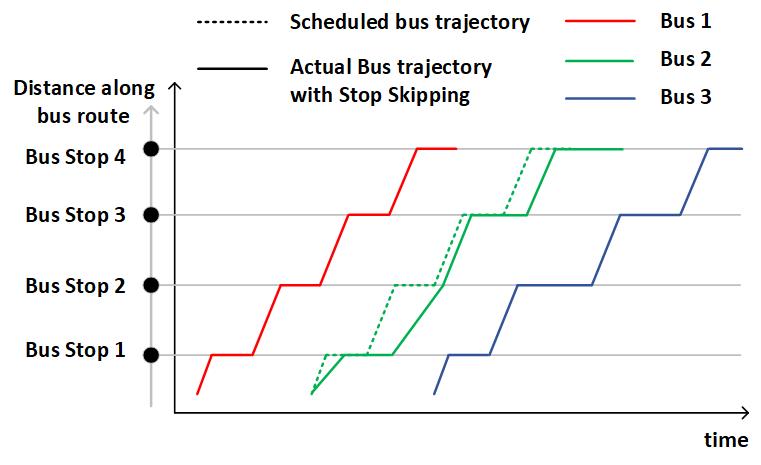}
        \caption{Stop-Skipping}\label{fig:skipconcept}
        \end{subfigure}
\\
\begin{subfigure}[b]{0.48\textwidth}\centering
        \includegraphics[width=\textwidth]{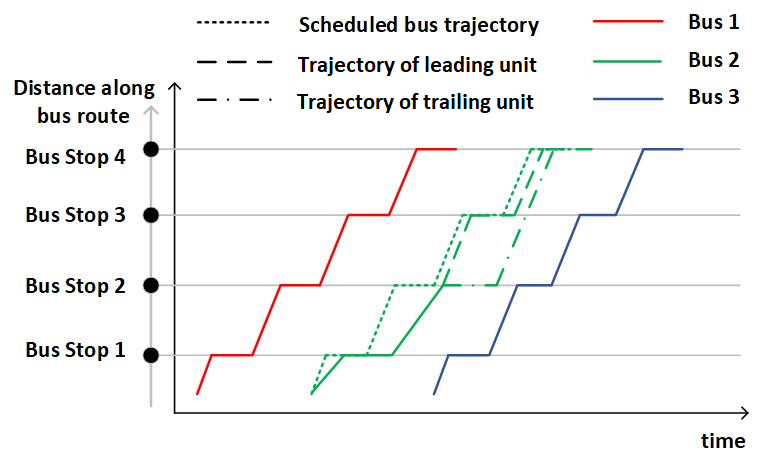}
        \caption{Bus-Splitting}\label{fig:splitconcept}
        \end{subfigure}
\caption{Time-space trajectories showing conceptual comparison of the proposed modular bus-splitting policy with the traditional stop-skipping policy and the base (no control) case when a bus experiences large headway deviations.}\label{fig:concept}
\end{figure}

We provide a simple example that shows how the introduction of modular bus units can improve the system reliability when a bus is late. To do so, we consider the following three cases: (i) a bus system consisting of conventional buses without any intervention (\textit{no control}); (ii) a bus system consisting of conventional buses wherein a bus skips stops when it is late (\textit{stop-skipping policy} which we consider the benchmark); and (iii) a bus system consisting of modular buses wherein a bus is split into modular units when it is late (our proposed \textit{bus-splitting policy}). These are shown in Figure~\ref{fig:concept}(a)-(c) respectively.

With the stop-skipping policy (Figure~\ref{fig:concept}(b)), when a bus falls behind schedule such that its time headway with the bus in front of it becomes larger than a given \textit{control threshold} times the target headway, the bus is directed to skip the subsequent stop (called the \textit{control stop}). The time saved by doing this allows it to reduce its headway and decrease its deviation from its scheduled trajectory. However, this comes at a cost to both the passengers waiting at the control stop (since they have to wait longer for the next bus) as well as the passengers on board who wanted to alight at the control stop (since they have to alight at the next stop and incur extra walking time back to the control stop). 

Using modular buses, the advantage offered by stop-skipping can be gained without incurring the associated cost. With the bus-splitting policy (Figure~\ref{fig:concept}(c)), when the headway grows larger than the control threshold times the target headway, the bus is directed to decouple (i.e. split) into two modular units en-route to the control stop. The first modular unit (called the \textit{leading unit}) skips the control stop, while the other (called the \textit{trailing unit}) stops at the control stop to serve boarding and alighting passengers. The leading unit has therefore reduced its headway just like the bus in the stop-skipping policy, while the passengers have not incurred the extra cost associated with the control stop being skipped. Note that this requires all the passengers who wish to alight at the control stop to move to the trailing unit before the units decouple. Various flavors of this policy are possible with regards to the subsequent operation of the units. For example, they may recouple at the next stop immediately downstream of the control stop or continue to operate separately until they happen to be close enough to recouple with minimal time lost. 

Based on the above conceptual example, we hypothesize that the proposed bus-splitting policy can reduce the passenger travel time while maintaining the same level of operation as the traditional stop-skipping policy in terms of headway deviation. The remainder of the paper serves to test this hypothesis.



\section{Model Formulation}\label{sec:model}

We choose a discrete macroscopic approach for our model, which does not track individual passengers through the system, but deals with passenger arrivals, boardings, alightings, and departures in a non-continuous aggregate manner. This choice makes our model more accurate than a continuous macroscopic model (which approximates passengers into flows) on one hand, while also having a much smaller computational load and data requirement than a microscopic model (which tracks individual passengers) on the other hand. We build our model dynamics in three stages: we begin by describing the basic system setting without any control mechanism, then modify it for the stop-skipping policy, and finally develop our proposed bus-splitting policy.

\subsection{Basic Setting Without Control}\label{sec:nocontrol}


We consider a bus line serving a cyclical route with $S$ stops. Passengers arrive at each stop $s \in \{1,S\}$ following a Poisson process with fixed rate $\rate{s}$ per second. The total hourly demand along the line is thus $\demand = 3600 \sum_{s=1}^{S} \rate{s}$. When a bus arrives at stop $s$, each passenger on board alights with probability $\prob{s}$. The fleet comprises $N$ buses with capacity $K$ each, and the target headway is $H$. The distance between consecutive stops $s$ and $s_+$ is denoted $\dist{s}$.\footnote{$s_+ = s \: (\text{mod } S) + 1$ refers to the next stop after stop $s$. This notation is used to loop back to stop $1$ after stop $S$ to emulate a cyclic route. The previous stop before stop $s$ is analogously denoted $s_-$.}
The average cruising speed of buses is $\busspeed$, so that the expected cruising time between stops $s$ and $s_+$ is given by $\excruise{s} = \nicefrac{\dist{s}}{\busspeed}$.  The boarding and alighting times per passenger are denoted $\boardpp$ and $\alightpp$ respectively, and the fixed extra time lost at each stop (due to acceleration, deceleration, opening and closing doors, etc) is denoted $E$. The buses continuously cycle around the route, restarting at stop $1$ after serving stop $S$. Therefore we use the notation $r \in \mathbb{N^+}$ to distinguish a bus run from the physical bus $n \in \{1,N\}$. A bus run $r$ is mapped to a physical bus $n$ by the relation
\begin{linenomath}\begin{equation}
    n = (r-1) \: (\text{mod } N) + 1,
    \label{run}
\end{equation}\end{linenomath}
indicating that the buses run in a fixed order without swapping positions. We will use the term ``bus $r$" rather than ``bus run $r$" where it is clear from the context.


The dynamics of the system are described below. For each bus arrival at a stop, the following quantities are calculated:

The cruising time of bus $r$ between stop $s$ and stop $s_+$ is given by
\begin{linenomath}\begin{equation}
    \cruise{r}{s} = \excruise{s} + \noise{r}{s},
    \label{cruise}
\end{equation}\end{linenomath}
where $\noise{r}{s}$ are independent identically distributed random error terms given by
\begin{linenomath}\begin{equation}
    \noise{r}{s} \sim \text{Gamma}(\cruiseshape,\cruisescale) - \cruiseshape\cruisescale.
    \label{noise}
\end{equation}\end{linenomath}
This results in $\noise{r}{s}$ being positively skewed with mean $0$, which is more suitable for cruising time than a normally distributed error term with no skew. Furthermore, it also allows $\noise{r}{s}$ to be bounded below but not above, which is 
also desirable since cruising time cannot be shorter than a certain minimum value, but may grow without bound. Note that in reality, the error terms would be positively correlated in time and space. However, they are commonly assumed to be independent for simplicity \cite{morales2019stochastic}.

The arrival time of bus $r$ at stop $s$ is given by
\begin{linenomath}\begin{equation}
    \arr{r}{s} = \max \left\{ \dep{r}{s_-} + \cruise{r}{s_-}, \; \dep{r-1}{s} \right\},
    \label{arr}
\end{equation}\end{linenomath}
where $\dep{r}{s_-}$ is the departure time of bus $r$ from the previous stop $s_-$ and $\dep{r-1}{s}$ is the departure time of the previous bus $r-1$ from stop $s$. The second term in the maximization ensures that a bus does not dock at a stop before the previous bus has departed \cite{sun2005real}. This is in line with the assumption that each bus stop has only one docking bay, which is common in practice. Furthermore, this maximization also ensures that buses do not overtake each other at any point, which is necessary to maintain the order of buses described above. Since two buses traveling on the same road segment at the same time experience the same traffic conditions, it is reasonable to assume that there will be no overtaking within the segments themselves.

The headway of bus $r$ with the downstream bus $r-1$ when arriving at stop $s$, called the arriving headway, is given by
\begin{linenomath}\begin{equation}
    \head{r}{s} = \arr{r}{s} - \arr{r-1}{s}.
    \label{head}
\end{equation}\end{linenomath}

The load on bus $r$ when it arrives at stop $s$ is given by
\begin{linenomath}\begin{equation}
    \load{r}{s} = \load{r}{s_-} + \board{r}{s_-} - \alight{r}{s_-},
    \label{load}
\end{equation}\end{linenomath}
where $\board{r}{s_-}$ and $\alight{r}{s_-}$ are the number of passengers boarding and alighting respectively at the previous stop. Given the alighting mechanism described earlier, $\alight{r}{s}$ is given by
\begin{linenomath}\begin{equation}
    \alight{r}{s} \sim \text{Binomial} ( \load{r}{s}, \prob{s} ).
    \label{alight}
\end{equation}\end{linenomath}

The number of passengers waiting to board bus $r$ at stop $s$ is given by
\begin{linenomath}\begin{equation}
    \wboard{r}{s} \sim \text{Poisson} ( \rate{s} \head{r}{s} ) + ( \wboard{r-1}{s} - \board{r-1}{s} ),
    \label{wboard}
\end{equation}\end{linenomath}
which includes both the passengers arriving during the current headway $\head{r}{s}$ and any left over passengers who could not board the previous bus. 
Note that this includes two assumptions for modeling simplicity: (i) passengers who arrive after a bus arrives at a stop are not served by the current bus, and (ii) there is no correlation between the number of passengers boarding and alighting at a stop. These simplifications do not have a significant effect on the dynamics \cite{morales2019stochastic}.

The actual number of passengers boarding the bus, which is limited by the available bus capacity, is given by
\begin{linenomath}\begin{equation}
    \board{r}{s} = \min \{ \wboard{r}{s} , \; K - (\load{r}{s} - \alight{r}{s}) \}.
    \label{board}
\end{equation}\end{linenomath}

The dwell time of bus $r$ at stop $s$ is then given by 
\begin{linenomath}\begin{equation}
    \dwell{r}{s} = \alightpp \alight{r}{s} + \boardpp \board{r}{s} + E,
    \label{dwell}
\end{equation}\end{linenomath}
assuming for simplicity that boarding and alighting occur sequentially. \footnote{Adopting a simultaneous boarding and alighting process instead of this would not affect the findings of the paper.}

The departure time of bus $r$ from stop $s$ is given by
\begin{linenomath}\begin{equation}
    \dep{r}{s} = \arr{r}{s} + \dwell{r}{s}.
    \label{dep}
\end{equation}\end{linenomath}

Finally, the headway of bus $r$ with the downstream bus $r-1$ when departing from stop $s$, called the departing headway, is given by
\begin{linenomath}\begin{equation}
    \dephead{r}{s} = \dep{r}{s} - \dep{r-1}{s}.
    \label{dephead}
\end{equation}\end{linenomath}
This quantity is not required for the basic system dynamics, but will be necessary for the control policies described later. 

The initial conditions that must be specified for the system are the loads $\load{r}{1}$ and starting times  $\arr{r}{1}$ at the first stop $s=1$ for the first cycle, i.e. for $1 \leq r \leq N$.

\subsection{Stop-Skipping}\label{sec:skipping}

We now introduce a simple stop-skipping policy to serve as a benchmark for our proposed bus-splitting policy described in the next subsection. Both the stop-skipping and bus-splitting policies are non-predictive (i.e. they make control decisions based only on currently available information), distributed (i.e. each bus makes its own decision without a centralized controller), and myopic (i.e. each bus only tries to minimize its own deviation from the target headway). Note that stop-skipping and bus-splitting policies can be made quite complex and sophisticated, taking into account the load on the current bus as well as upstream and downstream buses, the expected arrival time of the upstream bus, the importance of the stop, and several other factors.
However, for the sake of simplicity and to avoid strong assumptions about information availability, we consider very simple forms of these policies in this proof-of-concept paper.

In essence, our stop-skipping policy simply dictates that whenever a bus experiences a departing headway which is greater than a certain control threshold $\thresh$ times the target headway, it skips the next stop (called the control stop). There are two restrictions to this decision: (i) a bus cannot skip two consecutive stops (to prevent passengers from facing excessive walking time back to their desired alighting stop\footnote{We assume that all passengers who were unable to alight at their desired stop alight at the next stop and walk back to their desired stop before exiting the system.}), and (ii) a stop cannot be skipped by two consecutive buses (to prevent excessive waiting time of passengers at that stop). We use an indicator variable $\serve{r}{s}$ to specify whether bus $r$ serves stop $s$. The control decision is given by
\begin{linenomath}\begin{equation}
    \serve{r}{s_+} = 
    \begin{cases}
      0 & \text{if } \dephead{r}{s} > \thresh H, \serve{r}{s} = 1, \text{ and } \serve{r-1}{s_+} = 1, \\
      1 & \text{otherwise}.
    \end{cases}
    \label{skip}
\end{equation}\end{linenomath}

This policy requires some modifications to the basic system dynamics described earlier. Firstly, if bus $r$ skips stop $s$, the passengers who wished to alight at stop $s$ would not be able to do so (we call them \textit{residual passengers}). Therefore we introduce the notation $\walight{r}{s}$ for the number of passengers who wish to alight, to distinguish them from $\alight{r}{s}$, the number of passengers who actually alight. It is clear that 
\begin{linenomath}\begin{equation}
    \alight{r}{s} = \serve{r}{s} \walight{r}{s}.
    \label{alight2}
\end{equation}\end{linenomath}

Assuming that the residual passengers all alight at the next stop $s_+$ (which cannot be skipped), we get
\begin{linenomath}\begin{equation}
    \walight{r}{s} \sim \text{Binomial} ( \load{r}{s} - ( \walight{r}{s_-} - \alight{r}{s_-} ), \prob{s} ) + ( \walight{r}{s_-} - \alight{r}{s_-} ).
    \label{walight}
\end{equation}\end{linenomath}
Equations~\eqref{alight2} and \eqref{walight} replace equation~\eqref{alight} from the basic setting. Because the passengers on board when arriving at stop $s$ may now include residual passengers from the previous stop, the definition of the alighting probability $\prob{s}$ must be updated accordingly. We now say that each passenger who did not wish to alight at any stop before stop $s$ wishes to alight at stop $s$ with probability $\prob{s}$ (i.e. the definition now explicitly refers to non-residual passengers only).

Similarly, the number of passengers who board now also depends on whether or not bus $r$ serves stop $s$. Equation~\eqref{board} is replaced by 
\begin{linenomath}\begin{equation}
    \board{r}{s} = \serve{r}{s} \min \{ \wboard{r}{s} , \; K - (\load{r}{s} - \alight{r}{s}) \}.
    \label{board2}
\end{equation}\end{linenomath}

Since the fixed extra time $E$ is only incurred if the stop is served, the dwell time equation~\eqref{dwell} is replaced by
\begin{linenomath}\begin{equation}
    \dwell{r}{s} = \alightpp \alight{r}{s} + \boardpp \board{r}{s} + \serve{r}{s} E.
    \label{dwell2}
\end{equation}\end{linenomath}
Note that due to the inclusion of $\serve{r}{s}$ in equations~\eqref{alight2}, \eqref{board2} and \eqref{dwell2}, $\dwell{r}{s}$ is $0$ if the stop is skipped.

    

\subsection{Bus-Splitting}\label{sec:splitting}

We consider each modular bus to consist of two identical modular units coupled together. Each unit has capacity $\nicefrac{K}{2}$. The units can decouple (split) and recouple in the order of a few seconds. Our bus-splitting policy is analogous to the stop-skipping policy in its simplicity. It dictates that whenever a bus experiences a departing headway which is greater than $\thresh H$, its modular units decouple en-route to the next stop, denoted the control stop $s'$. The leading unit is denoted $r'$ and the trailing unit is denoted $r''$, with $r$ continuing to refer to the aggregate of the two units. The leading unit skips stop $s'$ and serves the subsequent stop $s'_+$. The trailing unit serves stop $s'$ and stops at stop $s'_+$ only to alight passengers and recouple with the leading unit (i.e. boarding is not allowed). This ensures that no passengers are forced to miss their desired alighting stop, meaning that there is no extra walking time incurred. When the decision to split is made, passengers who wish to alight at stop $s'$ are asked to move to the trailing unit, while those who wish to alight at stop $s'_+$ are asked to move to the leading unit. This exchange, as well as the decoupling, happens while the bus is traversing the segment between $s'_-$ and $s'$. The units reach $s'$ at the same time. The load is assumed to be equally split among the two units, which requires that no more than half of the passengers on board wish to alight at either stop $s'$ or $s'_+$.\footnote{This constraint can be relaxed with a slight alteration to the dynamics, which we avoid for simplicity.} 

The restriction that a stop cannot be skipped by two consecutive buses is not applicable for bus-splitting since no stop is skipped completely. Therefore, a stop may become a control stop for consecutive buses. However, the other restriction that two consecutive stops cannot both be control stops continues to apply because we assume that a decoupled modular unit cannot be split further. Once the units have recoupled at stop $s'_+$, they may decouple again immediately afterward. 
We use an indicator variable $\split{r}{s}$ to specify whether bus $r$ is split at stop $s$ (in which case the stop would be denoted $s'$). The control decision is given by
\begin{linenomath}\begin{equation}
    \split{r}{s_+} = 
    \begin{cases}
      1 & \text{if } \dephead{r}{s} > \thresh H \text{ and } \split{r}{s}=0, \\
      0 & \text{otherwise}.
    \end{cases}
    \label{split}
\end{equation}\end{linenomath}

The equations describing the dynamics of the stop-skipping setting continue to apply here, although they apply separately to the units $r'$ and $r''$ for the duration that they are decoupled (i.e. at stops $s'$ and $s'_+$), rather than to the aggregate entity $r$. During this time, unit $r'$ considers its predecessor for its calculations to be bus $r-1$ which is the previous bus downstream of the split bus $r$. Unit $r''$ considers its predecessor to be unit $r'$. The next bus $r+1$ upstream of the split bus $r$ considers its predecessor to be unit $r''$. 
Other additions and modifications to the dynamics are described below. 

As a result of the splitting decision $\split{r}{s'} = 1$, the indicator variable $\serve{r'}{s'}$ is set to $0$ while $\serve{r''}{s'_+}$ remains $1$ since unit $r''$ does stop at stop $s'_+$. The load on each unit when arriving at stop $s'$ is given by
\begin{linenomath}\begin{align}
    \load{r'}{s'} & = \left\lceil \frac{\load{r}{s'}}{2} \right\rceil, \\
    \load{r''}{s'} & = \left\lfloor \frac{\load{r}{s'}}{2} \right\rfloor, \\
    \load{r}{s'} & = \load{r''}{s'} + \load{r'}{s'}.
    \label{load3}
\end{align}\end{linenomath}

The number of passengers wishing to alight from each unit at stop $s'$ is given by
\begin{linenomath}\begin{align}
    \walight{r'}{s'} & = 0, \\ 
    \walight{r''}{s'} & \sim \min \left\{ \text{Binomial} ( \load{r}{s'}, \prob{s} ), \; \load{r''}{s'} \right\}.
    \label{walight3a}
\end{align}\end{linenomath}
This is because all passengers who were on board bus $r$ when departing from stop $s'_-$ who wished to alight at stop $s'$ (which we assumed to be capped at half of the total) were moved to unit $r''$. Similarly, the number of passengers wishing to alight from each unit at stop $s'$ is given by
\begin{linenomath}\begin{align}
    \walight{r'}{s'_+} & \sim \min \left\{ \text{Binomial} ( \load{r}{s'} - \walight{r''}{s'}, \prob{s} ), \; \load{r'}{s'} \right\}, \\
    \walight{r''}{s'_+} & \sim \text{Binomial} ( \board{r''}{s'}, \prob{s} ).
    \label{walight3b}
\end{align}\end{linenomath}

The former equation asserts that the passengers who could potentially wish to alight from unit $r'$ at stop $s'_+$ must not include those who wished to alight at stop $s'$. The latter equation asserts that the only passengers who could potentially wish to alight from unit $r''$ at stop $s'_+$ are those who boarded at the previous stop $s'$, i.e. those wishing to travel only one stop. This is because all passengers who were on board bus $r$ when departing from stop $s'_-$ who wished to alight at stop $s'_+$ were moved to unit $r'$.

The number of passengers wishing to board unit $r'$ at stops $s'$ and $s'_+$,  is calculated as usual based on its arriving headway. 
The number of passengers wishing to board unit $r''$ at stop $s'$ is given by

\begin{linenomath}\begin{equation}
    \wboard{r''}{s'} = \wboard{r'}{s'}.
    \label{wboard3}
\end{equation}\end{linenomath}

This is because unit $r''$ arrives at stop $s'$ at the same time as unit $r'$, which skips $s'$, leaving behind the passengers who had arrived up to that time. 

The number of passengers boarding the units is capped by the unit capacity $\nicefrac{K}{2}$ rather than the aggregate bus capacity $K$, which requires modifying equation~\eqref{board2}. Since unit $r''$ does not board passengers at stop ${s'_+}$ as described earlier, we set
\begin{linenomath}\begin{equation}
    \board{r''}{s'_+} = 0.
    \label{board3}
\end{equation}\end{linenomath}

Finally, to allow the units to recouple at stop $s'_+$, the departure time of the rejoined bus must be the later of the individually calculated ready-to-depart times of the two units, as described below.
\begin{linenomath}\begin{equation}
    \dep{r}{s'_+} = \max \left\{ \dep{r'}{s'_+}, \dep{r''}{s'_+} \right\}.
    \label{dep3}
\end{equation}\end{linenomath}
%


\section{Experiments}\label{sec:experiments}

\subsection{Experimental Setting}\label{sec:setting}

Note that our methodology places no assumptions on the variability of the segment lengths $\dist{s}$, arrival rates $\rate{s}$, and alighting probabilities $\prob{s}$ across stops. For our proof-of-concept experiments, however, we focus our attention on a quasi-homogeneous bus route in which all stops have approximately equal importance with respect to these three quantities. For each stop $s$, $\dist{s}$, $\rate{s}$ and $\prob{s}$ are drawn from normal distributions with means $\avgdist$, $\avgrate$, and $\avgprob$ respectively, and standard deviations equal to $10\%$ of the respective mean. This allows us to model a relatively symmetric system with some degree of heterogeneity between stops, rather than the unrealistic case of a completely homogeneous system. 

Under a uniform OD demand setting, the average number of stops traveled by a passenger is $\nicefrac{S}{2}$. In a cyclic route with probability $\avgprob$ of alighting at each stop, the number of stops traveled is given by a geometric distribution, which has mean $\nicefrac{1}{\avgprob}$. Therefore, we require $\avgprob = \nicefrac{2}{S}$.\footnote{Note that $\prob{s}$ only represents the probability of alighting when arriving at stop $s$; it does not represent the fraction of passengers whose desired destination is $s$ (and therefore $\sum \prob{s}$ does not need to be $1$).}

We select our initial conditions with the goal of starting the experiments with the system in (unstable) equilibrium. The load $\load{r}{1}$ on each bus when starting its first cycle (i.e. for $1 \leq r \leq N$ and stop $s = 1$) is equal to the expected average load, given by
\begin{linenomath}\begin{equation}
    L = \frac{S \avgrate H}{2}.
    \label{avgload}
\end{equation}\end{linenomath}

The corresponding starting time $\arr{r}{1}$ for each bus is given by $(r-1) H$, indicating that the buses are initially dispatched with 
headway $H$, with the first bus starting at time $0$.

We now describe how we determine the fleet size $N$ and target headway $H$. First, note that the expected cycle time of a bus is given by 
\begin{linenomath}\begin{equation}
    \cycle = \left( \avgcruise + (\alightpp + \boardpp) \avgrate H + E \right) S = NH,
    \label{cycle}
\end{equation}\end{linenomath}
where $\avgcruise = \nicefrac{\avgdist}{\busspeed}$ is the expected cruising time. Next, note that satisfying a uniform OD demand requires $K \geq L$. Using \eqref{avgload} and \eqref{cycle}, this can be rewritten in terms of $N$ as 
\begin{linenomath}\begin{equation}
    N \geq \frac{S \avgrate \cycle}{2K}.
    \label{Nineq}
\end{equation}\end{linenomath}

Eliminating dependence on $H$ using \eqref{cycle}, the minimum fleet size $\minfleet$ required is given by
\begin{linenomath}\begin{equation}
    \minfleet = (\alightpp + \boardpp) S \avgrate + \frac{(\avgcruise+E) S^2 \avgrate}{2K}.
    \label{minfleet}
\end{equation}\end{linenomath}

In order to account for the non-uniformity of the demand, and to reduce the expected load $L$ (which would be near capacity for $N = \lceil \minfleet \rceil$, leading to a precariously unstable system), we choose 
\begin{linenomath}\begin{equation}
    N = \lceil \fleetmult \minfleet \rceil,
    \label{fleetsize}
\end{equation}\end{linenomath}
where $\fleetmult>1$ is the fleet size multiplier. The target headway is then calculated by solving \eqref{cycle} for $H$. The parameter values we use for our experiments are shown in Table~\ref{tab:paras}.

\begin{table}[!ht]\centering
        \caption{Independent Parameter Values}\label{tab:paras}
        \begin{tabular}{l l l l}
                Input Parameter & Notation & Units & Value \\\hline
                Number of stops & $S$ & - & 20 \\
                Average spacing between stops & $\avgdist$ & m & 400 \cite{tirachini2014economics} \\
                Bus capacity & $K$ & pax & 80 \\
                Unit capacity & $\nicefrac{K}{2}$ & pax & 40 \\
                Bus cruising speed & $\busspeed$ & km/h & 20 \cite{kim2003performance} \\
                Walking speed & $\walkspeed$ & km/h & 4.5 \cite{chandra2013speed} \\
                Fixed time lost per stop & $E$ & s & 20 \cite{kfh2013transit} \\
                Boarding time per passenger & $\boardpp$ & s/pax & 4 \cite{kfh2013transit} \\
                Alighting time per passenger & $\alightpp$ & s/pax & 3 \cite{kfh2013transit} \\
                Waiting time weight factor & $\waitmult$ & - & 2.1 \cite{wardman2004public} \\
                Walking time weight factor & $\walkmult$ & - & 2.2 \cite{wardman2004public} \\
                Fleet size factor & $\fleetmult$ & - & 1.5 \\
                \hline
        \end{tabular}
\end{table}

We evaluate the three policies, no control, stop-skipping, and bus-splitting, with respect to the average travel cost $\avgcost$ per passenger, which is a weighted sum of the waiting time $\waittime$, in-vehicle time $\vehtime$, and walking time $\walktime$. Following convention, the weight attached to each component is the value of a unit of that component as a multiple of the value of a unit of in-vehicle time. By definition, this means that the weight of in-vehicle time is $1$. The weights of the other two components are shown in Table~\ref{tab:paras}. The equation for the average travel cost is
\begin{linenomath}\begin{equation}
    \avgcost = \waitmult \waittime + \vehtime + \walkmult \walktime.
    \label{avgcost}
\end{equation}\end{linenomath}

It is also useful to determine how much additional overhead is caused by bus bunching compared to the expected travel cost in the absence of stochasticity. The expected (non-stochastic) travel cost, denoted $\expcost$, is given by
\begin{linenomath}\begin{equation}
    \expcost = \frac{(\waitmult + N) H}{2}.
    \label{expcost}
\end{equation}\end{linenomath}

The bunching overhead $\overhead$ is then the difference between the average travel cost and the expected travel cost as a percentage of the latter:
\begin{linenomath}\begin{equation}
    \overhead = \frac{\avgcost-\expcost}{\expcost} 100\%.
    \label{overhead}
\end{equation}\end{linenomath}

\subsection{Simulation}\label{sec:simulation}

We simulate our system in Python using a discrete event simulation in which updates are processed each time a bus arrives at a stop. In order to ensure that the system is in a stable state, we allow the whole fleet to make two complete cycles before beginning our \textit{evaluation period}, which lasts for the next one hour. All our results are based on the evaluation period only. Since we do not track individual passengers, the average waiting time, in-vehicle time, and walking time are calculated using the area between the respective cumulative functions divided by the number of corresponding events.\footnote{For example, the average in-vehicle time is given by the area between the cumulative boarding and alighting functions, divided by the average of the number of boardings and alightings.} This method of calculating average times is approximate rather than exact, but the error is insignificant when the system is in a stable state.


\section{Results}\label{sec:results}

\subsection{Sample Comparison}\label{sec:sample}

\begin{figure}[!b]\centering
        \begin{subfigure}[b]{0.45\textwidth}\centering
        \includegraphics[width=\textwidth]{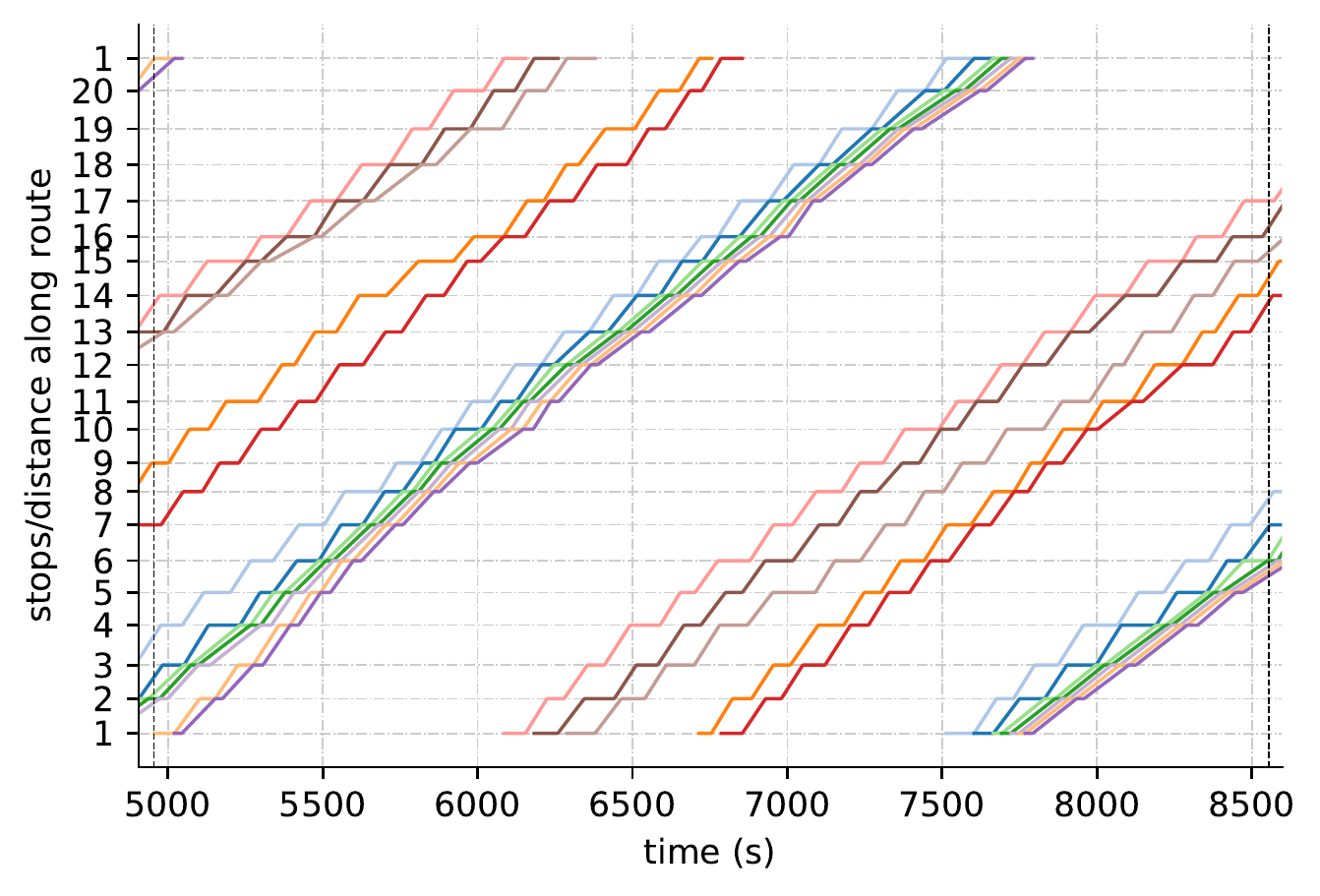}
        \caption{No control}\label{fig:nocontroltraj}
        \end{subfigure}
        \begin{subfigure}[b]{0.45\textwidth}\centering
        \includegraphics[width=\textwidth]{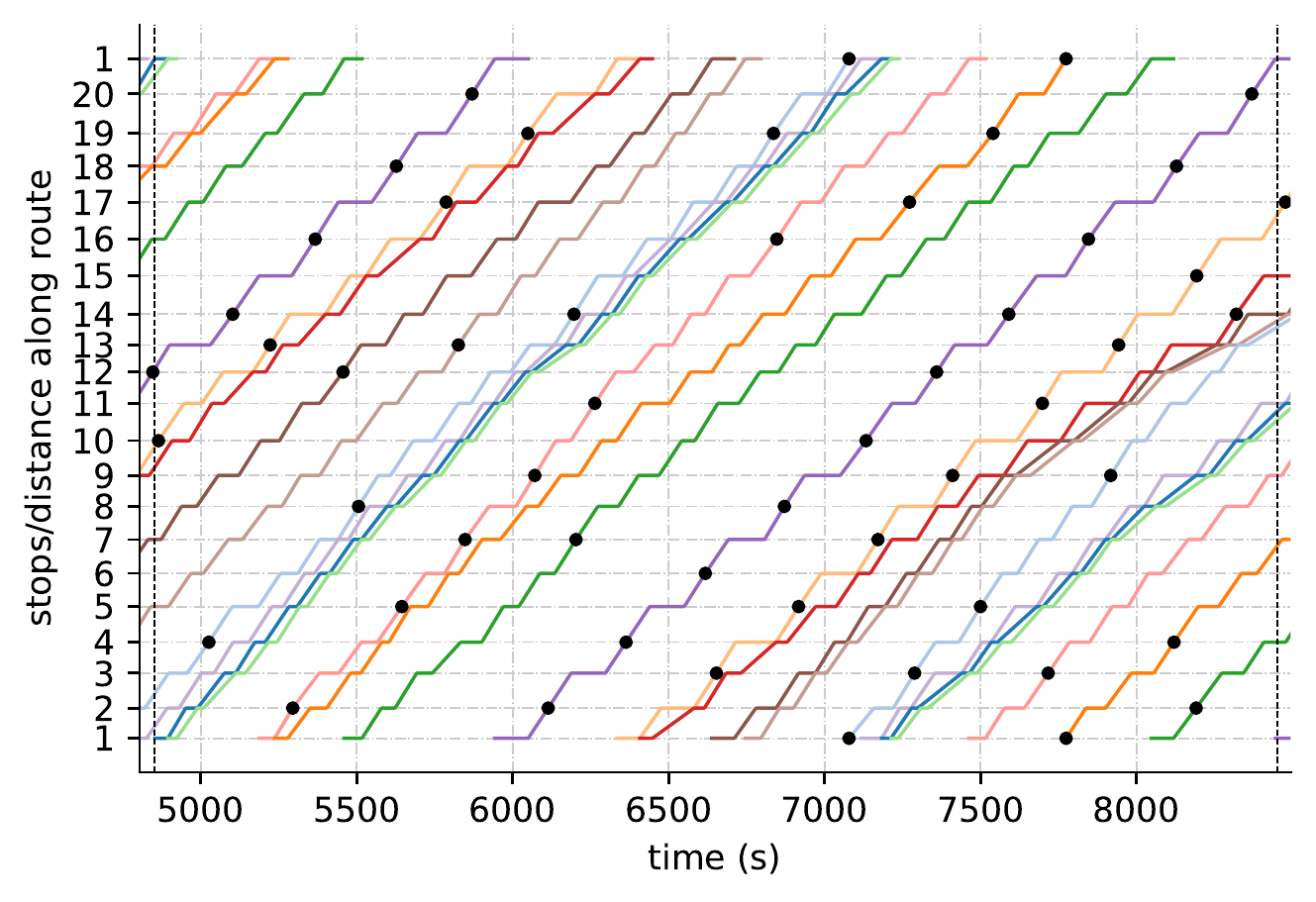}
        \caption{Stop-Skipping}\label{fig:skiptraj}
        \end{subfigure}
\\
\begin{subfigure}[b]{0.45\textwidth}\centering
        \includegraphics[width=\textwidth]{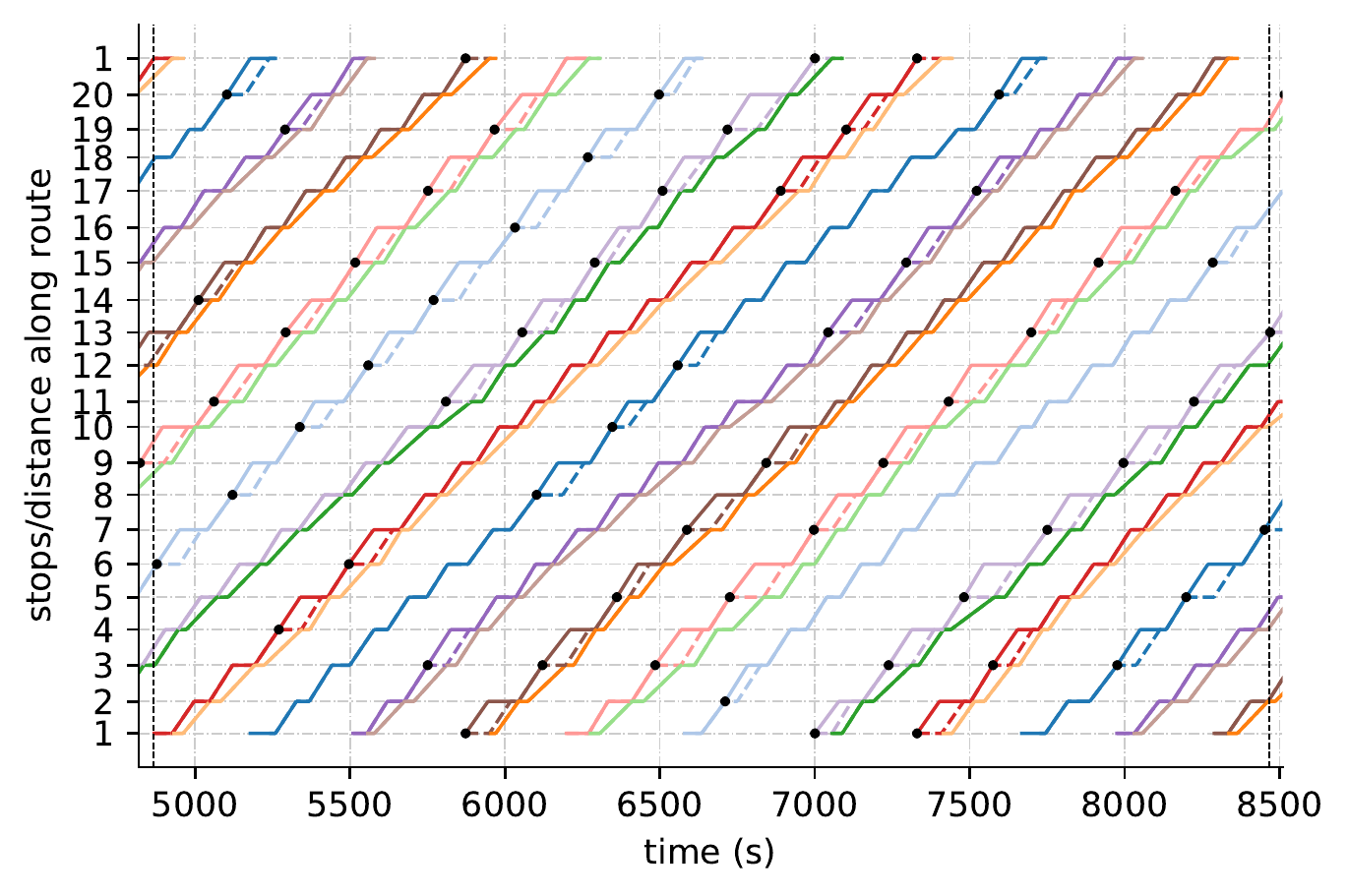}
        \caption{Bus-Splitting}\label{fig:splittraj}
        \end{subfigure}
\caption{Bus trajectories during the evaluation period from a representative instance under each policy. Each color represents a particular bus making multiple cycles. The black dots represent stops where the control action is triggered. The dashed lines in (c) represent the trailing units. Parameter values are taken from Table~\ref{tab:paras}. The hourly demand is $\demand=1500$, the fleet size is $N=12$ and the target headway is $H=203$s ($3.4$min).}\label{fig:traj}
\end{figure}

Figure~\ref{fig:traj} presents time-space plots showing the bus trajectories during the evaluation period for a representative instance under each of the three policies. Without any form of control (Figure~\ref{fig:traj}(a)), the system deteriorates into severe bus bunching, with gaps many times larger than the target headway and bunches containing up to half the fleet. Under the stop-skipping policy (Figure~\ref{fig:traj}(b)), the upper limit of the headways is much smaller, preventing the system from collapsing completely, but there are still large bunches and significant headway variability. The bus-splitting policy (Figure~\ref{fig:traj}(c)) is also unable to completely eradicate bus bunching (which is expected due to its simplicity)
, but results in smaller bunches which maintain relatively even headways from each other, giving the system some stability. The trajectory dynamics seen here suggest that a more sophisticated bus-splitting policy which considers the trailing headway with the upstream bus in addition to the (currently considered) leading headway with the downstream bus could be more effective at alleviating bunching. This will be discussed further in the conclusion.

\begin{figure}[!b]\centering
        \begin{subfigure}[b]{0.45\textwidth}\centering
        \includegraphics[width=\textwidth]{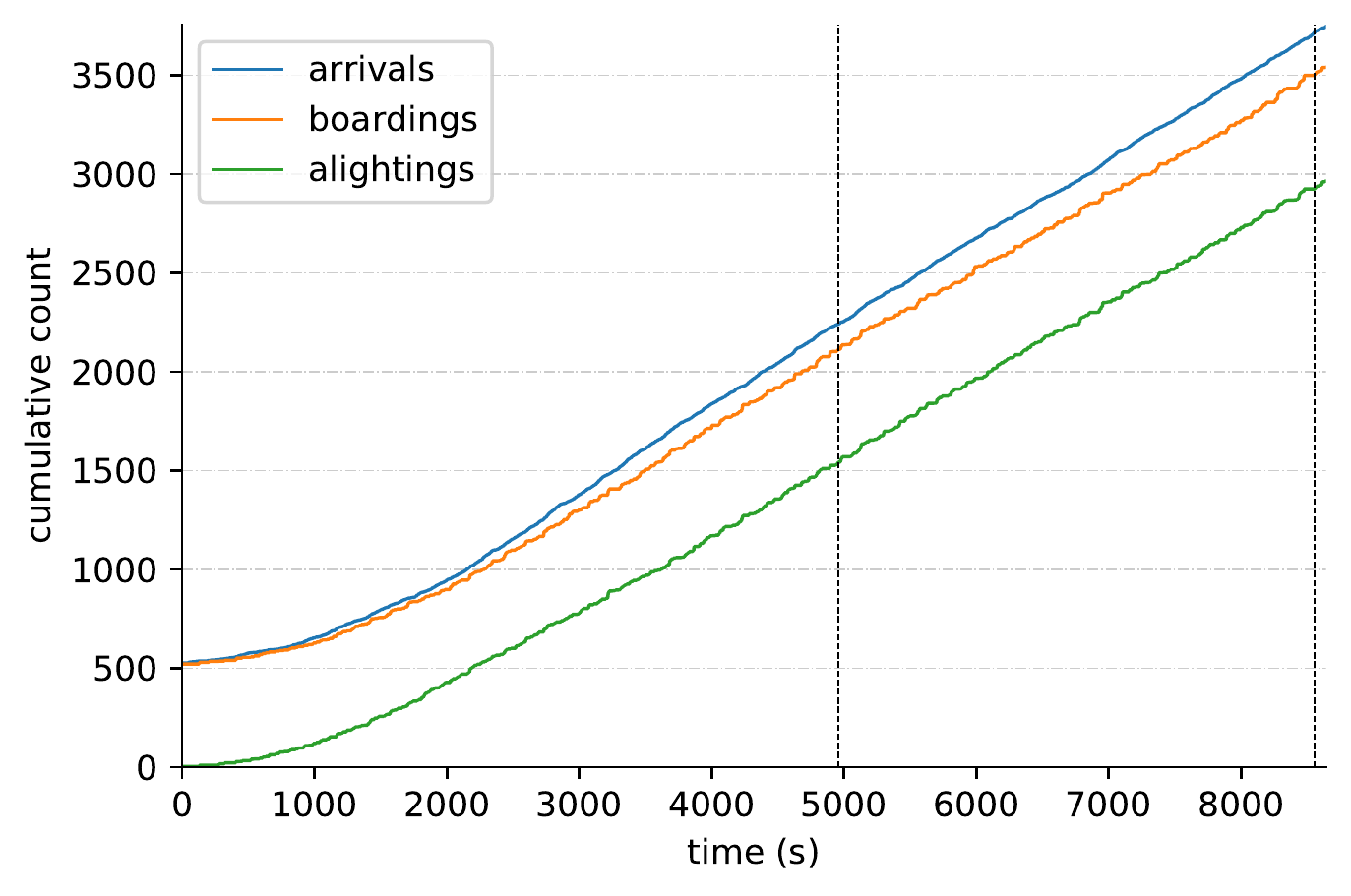}
        \caption{No control}\label{fig:nocontrolcumulative}
        \end{subfigure}
        \begin{subfigure}[b]{0.45\textwidth}\centering
        \includegraphics[width=\textwidth]{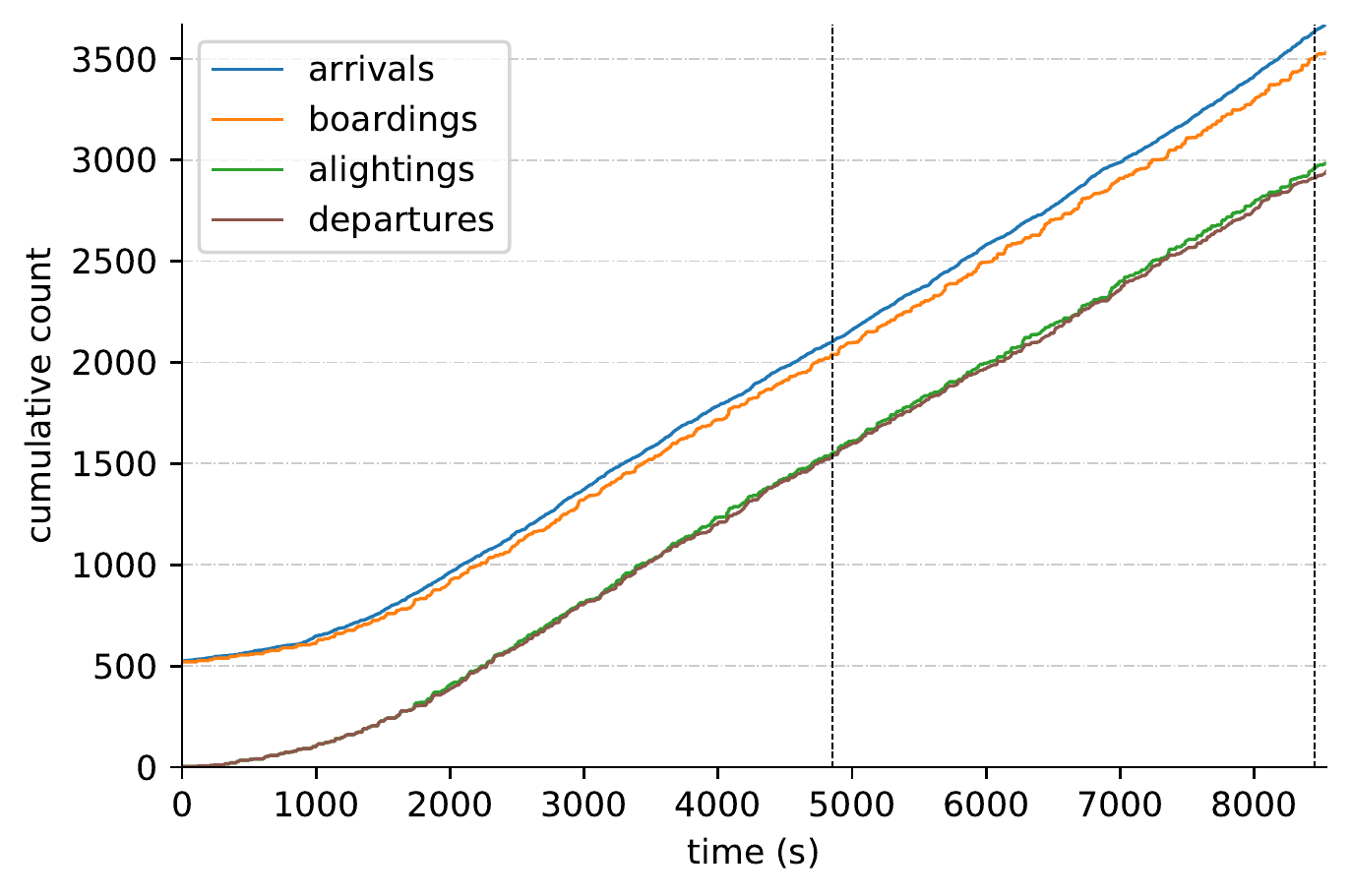}
        \caption{Stop-Skipping}\label{fig:skipcumulative}
        \end{subfigure}
\\
        \begin{subfigure}[b]{0.45\textwidth}\centering
        \includegraphics[width=\textwidth]{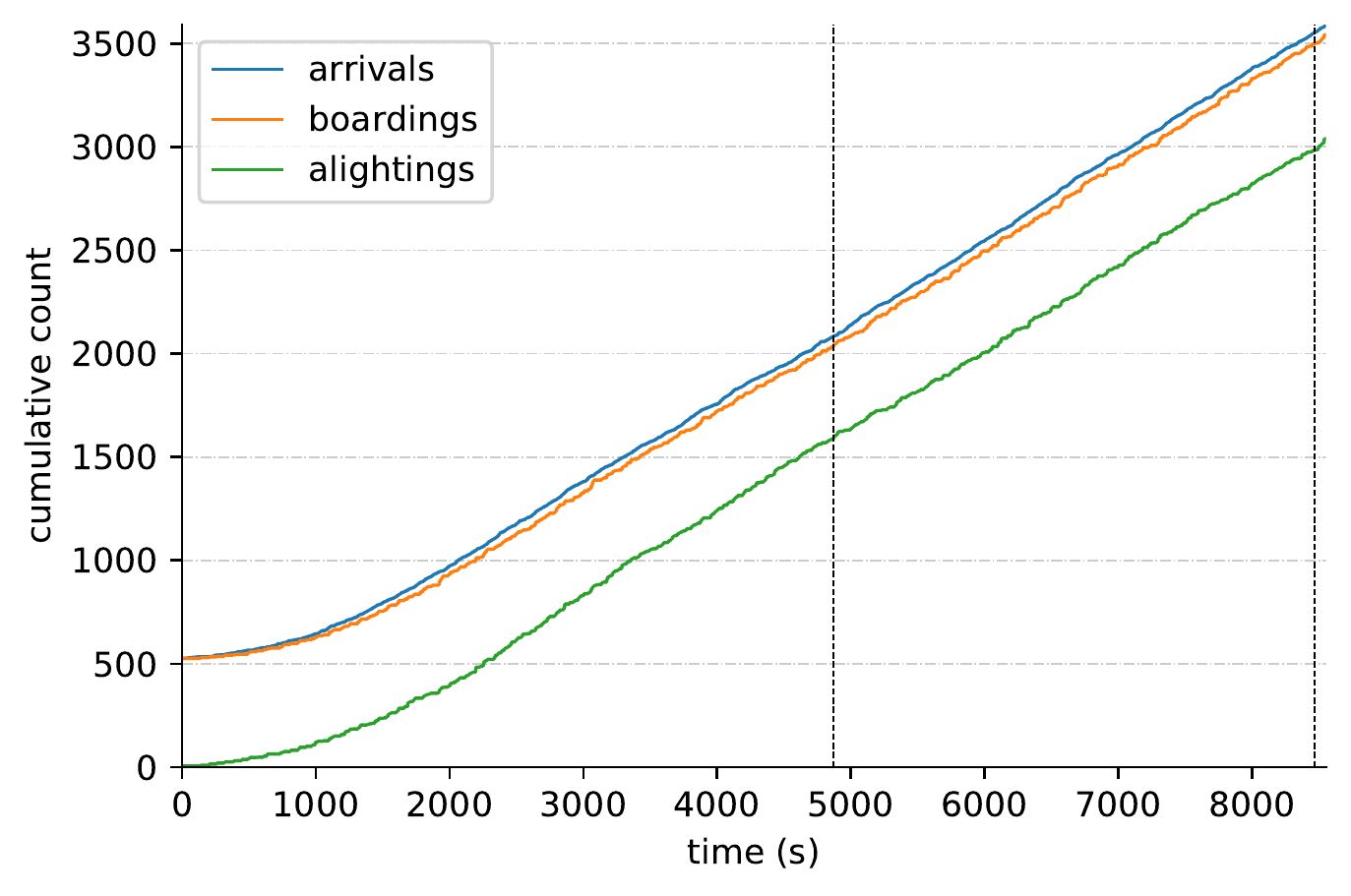}
        \caption{Bus-Splitting}\label{fig:splitcumulative}
        \end{subfigure}
        \begin{subfigure}[b]{0.45\textwidth}\centering
        \includegraphics[width=\textwidth]{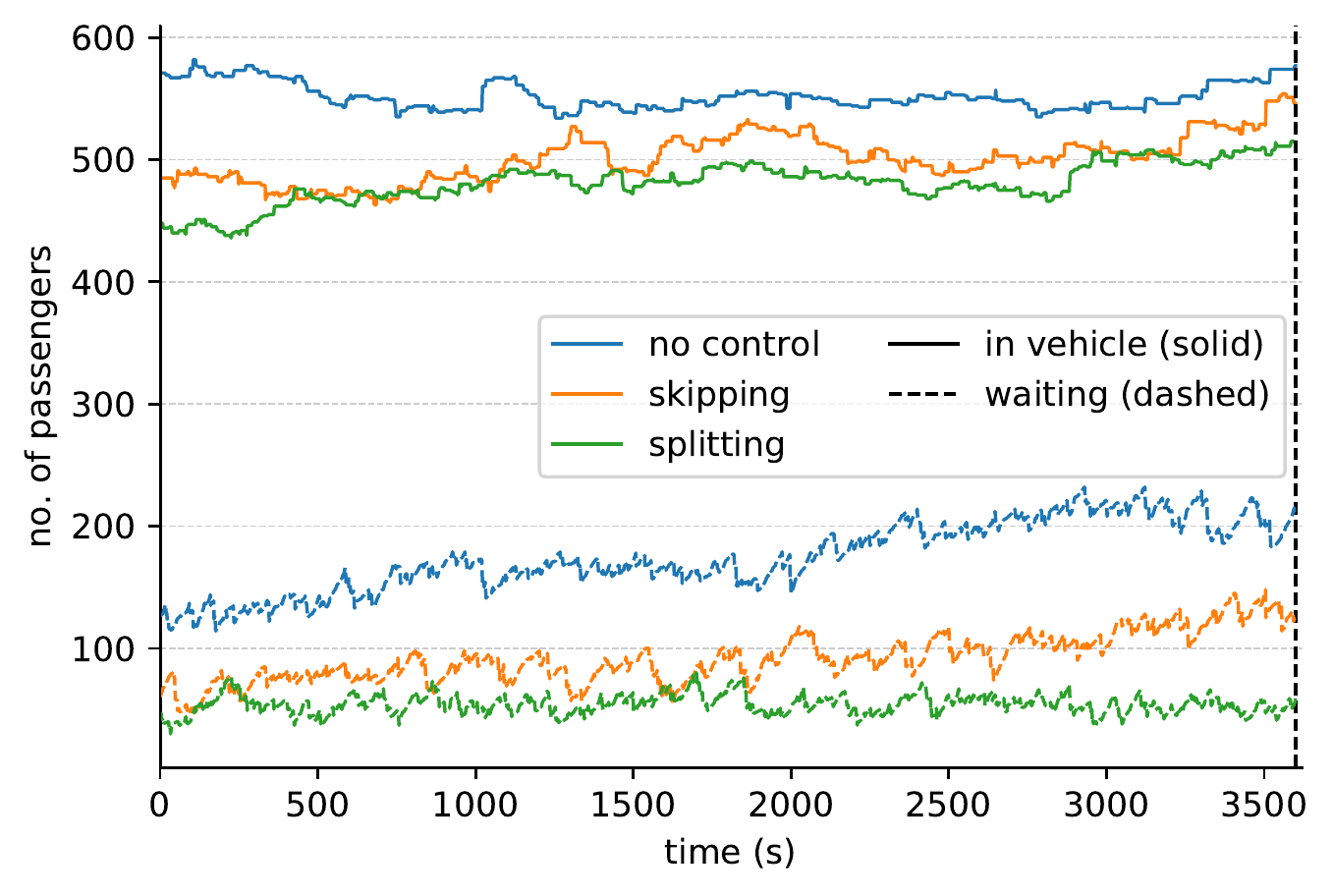}
        \caption{Comparison}\label{fig:waitingandinvehicle}
        \end{subfigure}
\caption{(a)-(c): cumulative passenger arrivals, boardings, alightings, and departures for the instances shown in Figure~\ref{fig:traj}(a)-(c) respectively. (d): number of passengers waiting to board and number of passengers on board during the evaluation period for each of the instances in (a)-(c). The vertical lines depict the beginning and end of the evaluation period.}\label{fig:cumulative}
\end{figure}

Figures~\ref{fig:cumulative}(a)-(c) show the cumulative passenger arrivals, boardings, alightings, and departures for the instances shown in Figures~\ref{fig:traj}(a)-(c) respectively. Without control, the effective capacity of the bus service is severely diminished, and the system is oversaturated as a result. This can be seen from the divergence between the cumulative arrivals and boardings. This effect is also present to a lesser extent under stop-skipping. Under bus-splitting, however, the effective capacity is sufficient to cope with the demand, and the system is stable. Recall that unlike the other policies, stop-skipping results in passengers whose desired alighting stop was skipped. The extra cumulative line in Figure~\ref{fig:cumulative}(b) represents their departure from the system after completing their walk back to that stop. Figure~\ref{fig:cumulative}(d) shows the number of passengers waiting to board a bus and the number of passengers on board a bus over the evaluation period for each of the instances in Figures~\ref{fig:traj}(a)-(c). The former represents the difference between the cumulative arrivals and boardings in the earlier figures, while the latter represents the difference between the cumulative boardings and alightings. The number of passengers waiting to board is consistently lowest under the bus-splitting policy, which suggests that it should have the lowest average waiting time. Similarly, the number of passengers on board is also lower under this policy, which suggests that it should also have a lower average in-vehicle time. The growing number of waiting passengers under the other two policies confirms that the system is oversaturated in both cases and will continue to deteriorate further, albeit more severely for the no control case.

\begin{table}[!b]\centering
        \caption{Sample Performance Comparison}\label{tab:metrics}
        \begin{tabular}{l l l l l l}
                Metric                          & Notation      & Units     & No Control & Stop-Skipping & Bus-Splitting \\\hline
                Average waiting time            & $\waittime$   & min       & 7.3       & 3.7   & 2.2   \\
                Average in-vehicle time         & $\vehtime$    & min       & 23.9      & 20.9  & 20.2  \\
                Average walking time            & $\walktime$   & min       & -         & 1.1   & -     \\
                Average travel cost             & $\avgcost$    & min       & 39.3      & 31.0  & 24.9  \\
                Bunching overhead               & $\overhead$   & \%        & 65.1      & 30.3  & 4.6   \\
                MAPE of headway                 & -             & \%        & 118.1     & 64.0  & 68.8  \\
                Average cycle length            & -             & min       & 47.0      & 40.6  & 40.2  \\
                Average load                    & -             & pax/bus   & 46.2      & 41.3  & 39.9  \\
                Fraction of full buses          & -             & -         & 0.29      & 0.02  & 0.00  \\
                \hline
        \end{tabular}
\end{table}

Table~\ref{tab:metrics} shows a comparison of several performance metrics for the instances shown in Figures~\ref{fig:traj}. No control is universally outperformed by stop-skipping, which is in turn outperformed by bus-splitting in nearly every metric. Both stop-skipping and bus-splitting drastically reduce the average passenger waiting time, which is expected since it is dependent on the headway variability. The $6$\textsuperscript{th} row of the table shows that both policies reduce this variability, quantified by the mean absolute percentage error (MAPE) with reference to the target headway $H$. The policies also reduce the average in-vehicle time, but this effect is smaller in percentage terms than on the average waiting time, since the passengers still traverse the same average distance and therefore face the same average cruising time. 
The average load and cycle time are also reduced by the two policies, indicating that they should be able to cope with a higher demand than without control, and would potentially require a smaller fleet size. This is reinforced by the next metric, which shows that without control at the current demand level, more than a quarter of the arrivals at stops are by buses which are completely full, whereas this fraction is nearly $0$ for both policies. 
The key observation from these metrics is that the bus-splitting policy results in a much smaller bunching overhead and travel cost compared to the stop-skipping policy, and that this is achieved by reducing each of the travel time components: waiting time, in-vehicle time, and walking time.

\subsection{Policy Robustness}\label{sec:sensitivity}

We now present results for a wide range of values of the control threshold $\thresh$ and demand $\demand$ in order to evaluate the effectiveness of the policies under different conditions. A smaller value of $\thresh$ results in a more aggressive (or proactive) policy, with the control action triggered even if headway is slightly longer than the target headway $H$. A larger value results in a more conservative policy, only triggering the control action if the headway becomes much longer than $H$. For each value of $\demand$, 
the values of $H$ and the fleet size $N$ 
are shown in Table~\ref{tab:rangevals}. Under all these settings, the target cycle time is $\cycle=40.5$ min and the target average load is $L=42$ passengers per bus.

\begin{table}[!ht]\centering
        \caption{Fleet Size and Target Headway for Each Demand Level}\label{tab:rangevals}
        \begin{tabular}{l l l}
                Demand & Fleet Size & Target Headway \\
                $\demand$ (pax/hr) & $N$ & $H$ (min) \\\hline
                250 & 2 & 20.3 \\
                500 & 4 & 10.1 \\
                750 & 6 & 6.8 \\
                1000 & 8 & 5.1 \\
                1250 & 10 & 4.1 \\
                1500 & 12 & 3.4 \\
                1750 & 14 & 2.9 \\
                2000 & 16 & 2.5 \\
                2250 & 18 & 2.3 \\
                2500 & 20 & 2.0 \\
                \hline
        \end{tabular}
\end{table}

We perform $500$ iterations with each set of parameter values and present the average results. Due to the extensive stochasticity inherent in the system, the results often have relatively high standard deviations, which are shown using error bars. 


\begin{figure}[!b]\centering
        \begin{subfigure}[b]{0.45\textwidth}\centering
        \includegraphics[width=\textwidth]{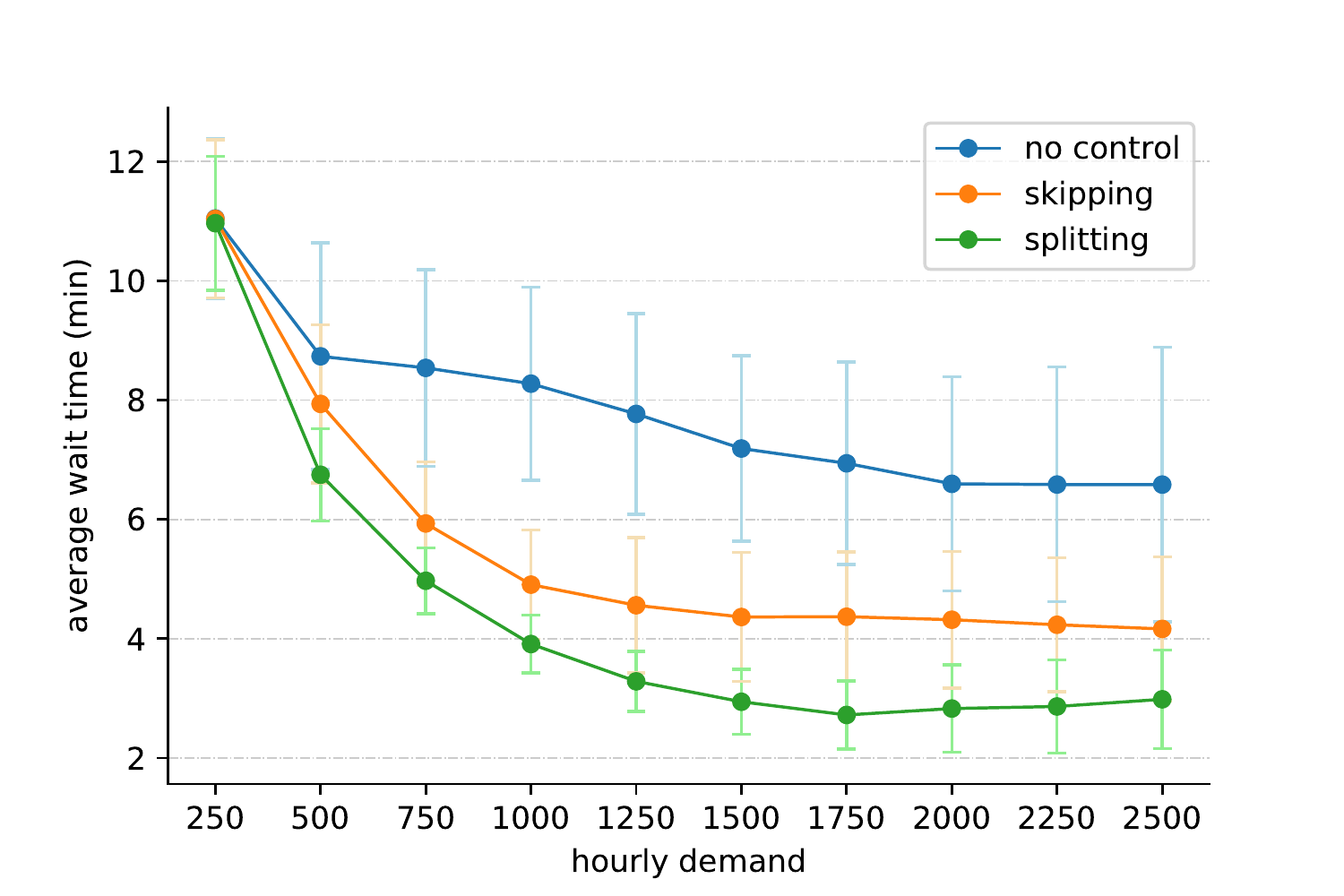}
        \caption{average waiting time}\label{fig:wait-thresh-15}
        \end{subfigure}
        \begin{subfigure}[b]{0.45\textwidth}\centering
        \includegraphics[width=\textwidth]{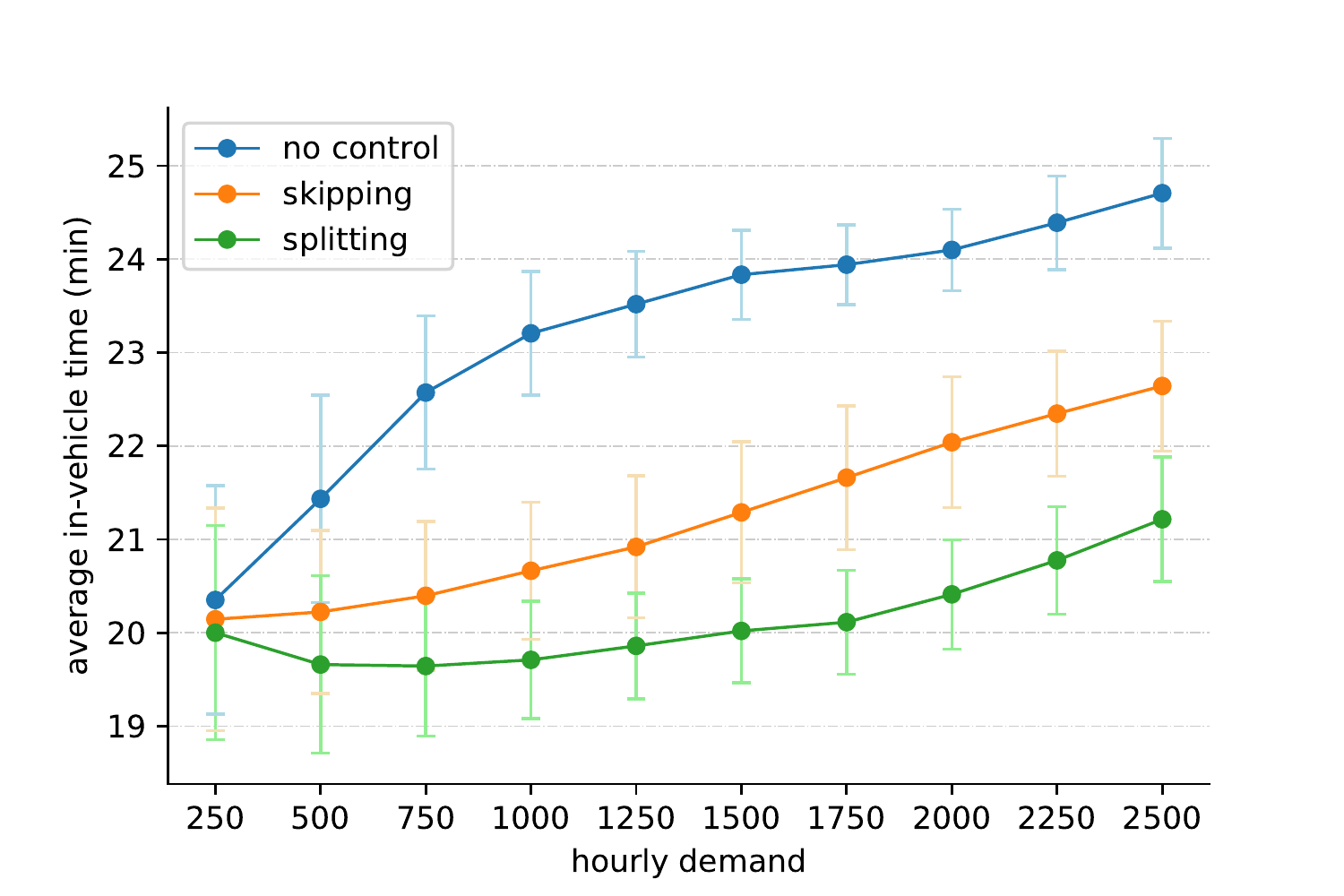}
        \caption{average in-vehicle time}\label{fig:in-vehicle-thresh-15}
        \end{subfigure}
\\       
        \begin{subfigure}[b]{0.45\textwidth}\centering
        \includegraphics[width=\textwidth]{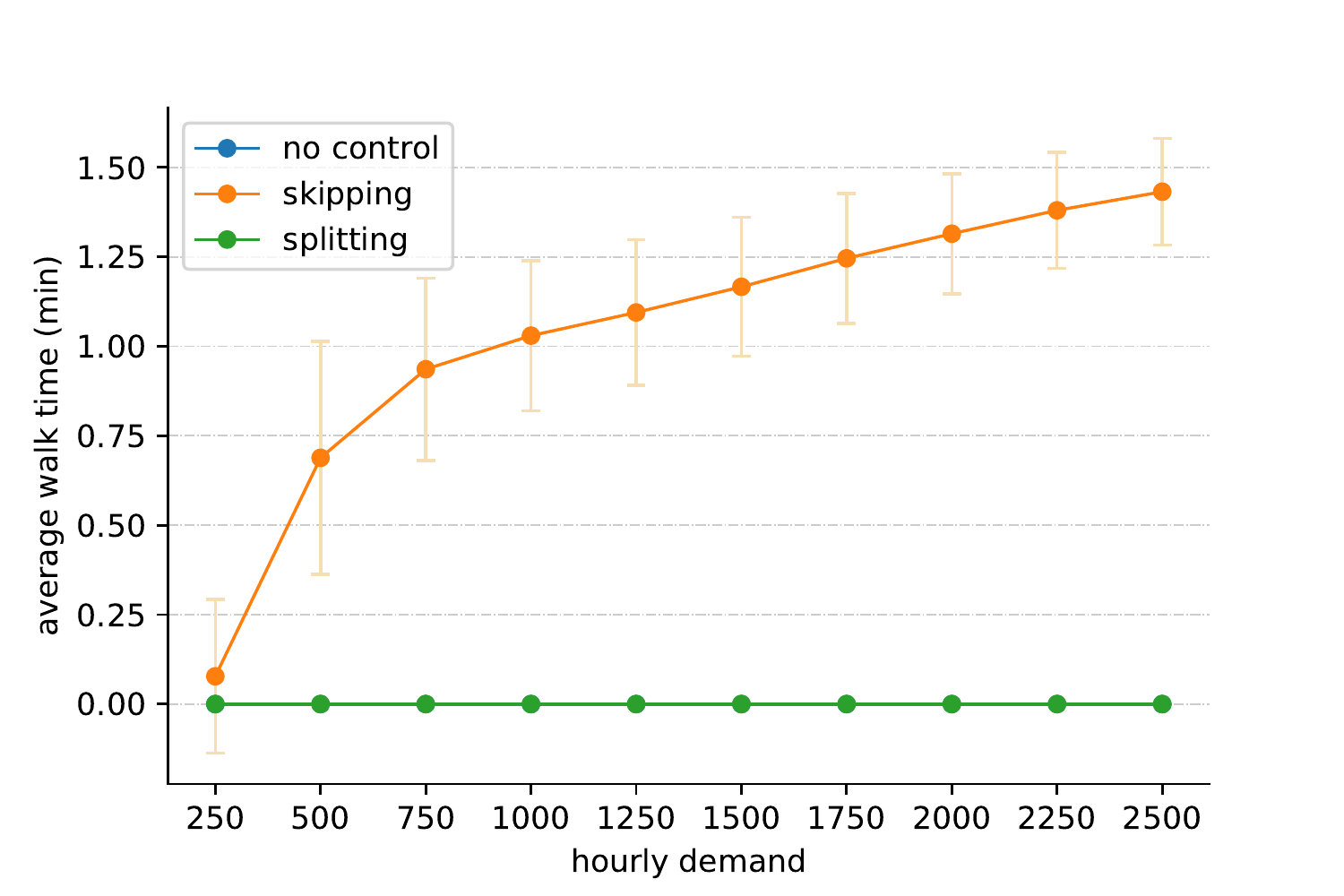}
        \caption{average walking time}\label{fig:walk-thresh-15}
        \end{subfigure}
        \begin{subfigure}[b]{0.45\textwidth}\centering
        \includegraphics[width=\textwidth]{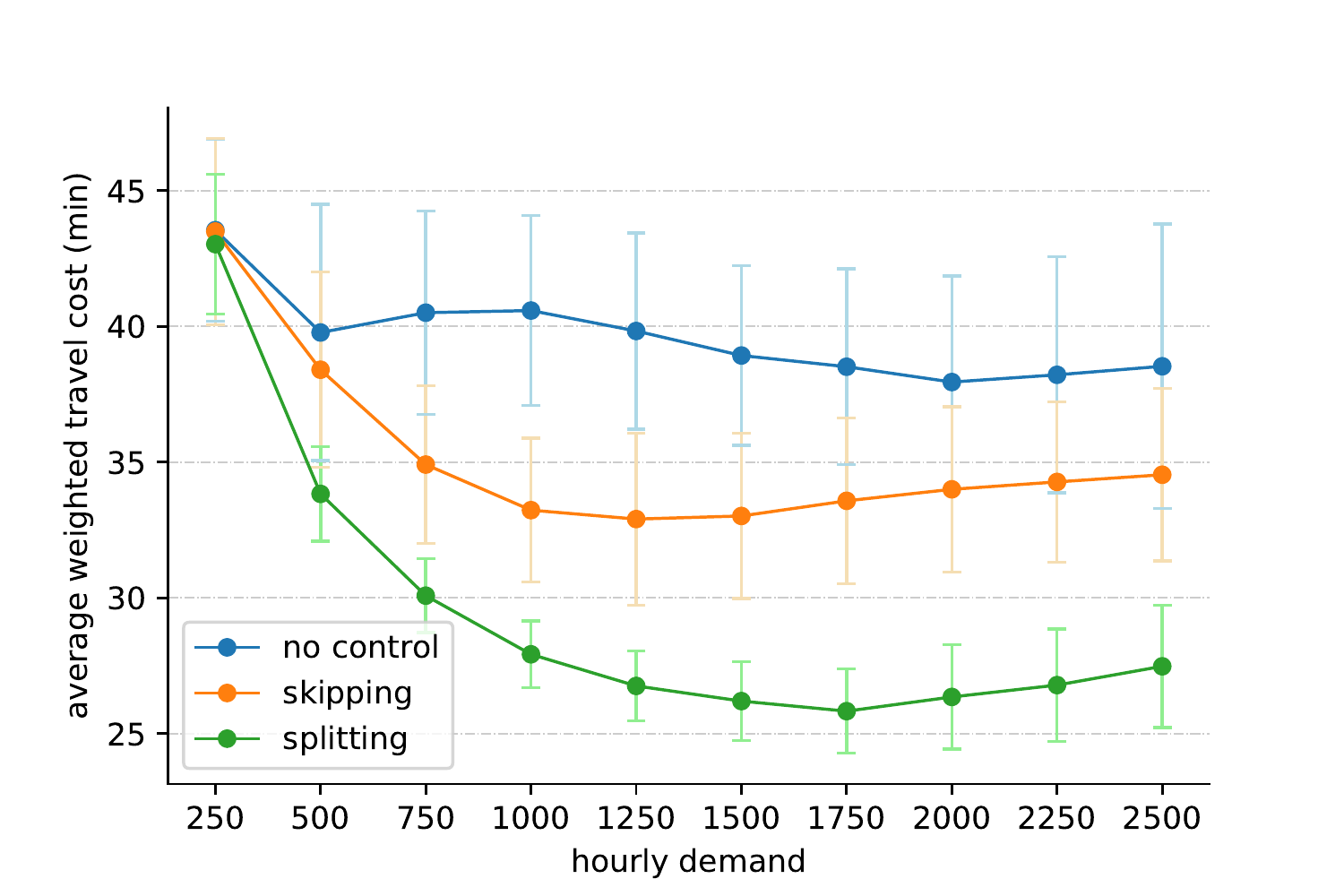}
        \caption{average weighted travel cost}\label{fig:weighted-thresh-15}
        \end{subfigure}
\caption{Average weighted travel cost $\avgcost$ (d) and each of its components (a)-(c) under a range of demand values $\demand$. The control threshold is $\thresh=1.5$. Error bars show standard deviation.}\label{fig:costs}
\end{figure}

Figure~\ref{fig:costs} shows how each of the travel time components and their weighted sum, the average travel cost $\avgcost$, vary with the demand $\demand$. Figure~\ref{fig:costs}(a) shows that the average waiting time decreases with increasing demand, particularly in the low demand range. This is expected since the fleet size is adjusted as the demand increases (to maintain the same average load), resulting in more frequent buses (i.e. smaller target headway $H$).\footnote{Note that for longer headways, the assumption of Poisson passenger arrivals ceases to be realistic, as passengers tend to time their arrival according to the scheduled bus arrival time. This means that for very low demand levels, the average waiting time will be lower in reality than seen here.} Once the demand is in the higher range, the waiting time becomes stable since further reductions in $H$ are minor, and are offset by the effect of increasing bus bunching.

In contrast, Figure~\ref{fig:costs}(b) shows that the average in-vehicle time increases with the demand. This is purely due to the effects of bus bunching, since the demand has no effect on the average distance traveled. Without control, the in-vehicle time rises steeply even in the low demand range, whereas under bus-splitting, it rises much more gradually until the demand becomes much higher. This indicates that bus-splitting is able to prevent much of the effects of bus bunching until the demand becomes too high to cope. Even under high demand, bus-splitting maintains a significantly lower in-vehicle time than stop-skipping, partially because the latter imposes additional in-vehicle time on those passengers whose intended alighting stop is skipped. 

Figure~\ref{fig:costs}(c) shows that the only policy that imposes any walking time is stop-skipping. Walking time rises steeply in the low demand range, as more and more stops are skipped in an attempt to reduce headway variability and prevent bus bunching. As demand increases further, the increase in walking time becomes more gradual, indicating that there are fewer remaining stops available for skipping under the restrictions of the policy, i.e. the execution of the control action is becoming saturated.

Finally, Figure~\ref{fig:costs}(d) shows the weighted sum of these three components, the average travel cost $\avgcost$. In the low demand range, the decrease in the waiting time is dominant, and $\avgcost$ decreases with increasing demand. The opposite is true in the high demand range, where the increase in the in-vehicle time is more significant. The bus-splitting policy maintains the decrease in $\avgcost$ longer than the stop-skipping policy, re-affirming that it is able to delay the effects of bus bunching until a higher demand level. Furthermore, the bus-splitting policy results in lower variation in the average travel cost for any given demand. This indicates a more reliable travel experience, which holds great importance for passengers' mode choice decisions and utility \cite{nesheli2015}.


\begin{figure}[!b]\centering
        \begin{subfigure}[b]{0.45\textwidth}\centering
        \includegraphics[width=\textwidth]{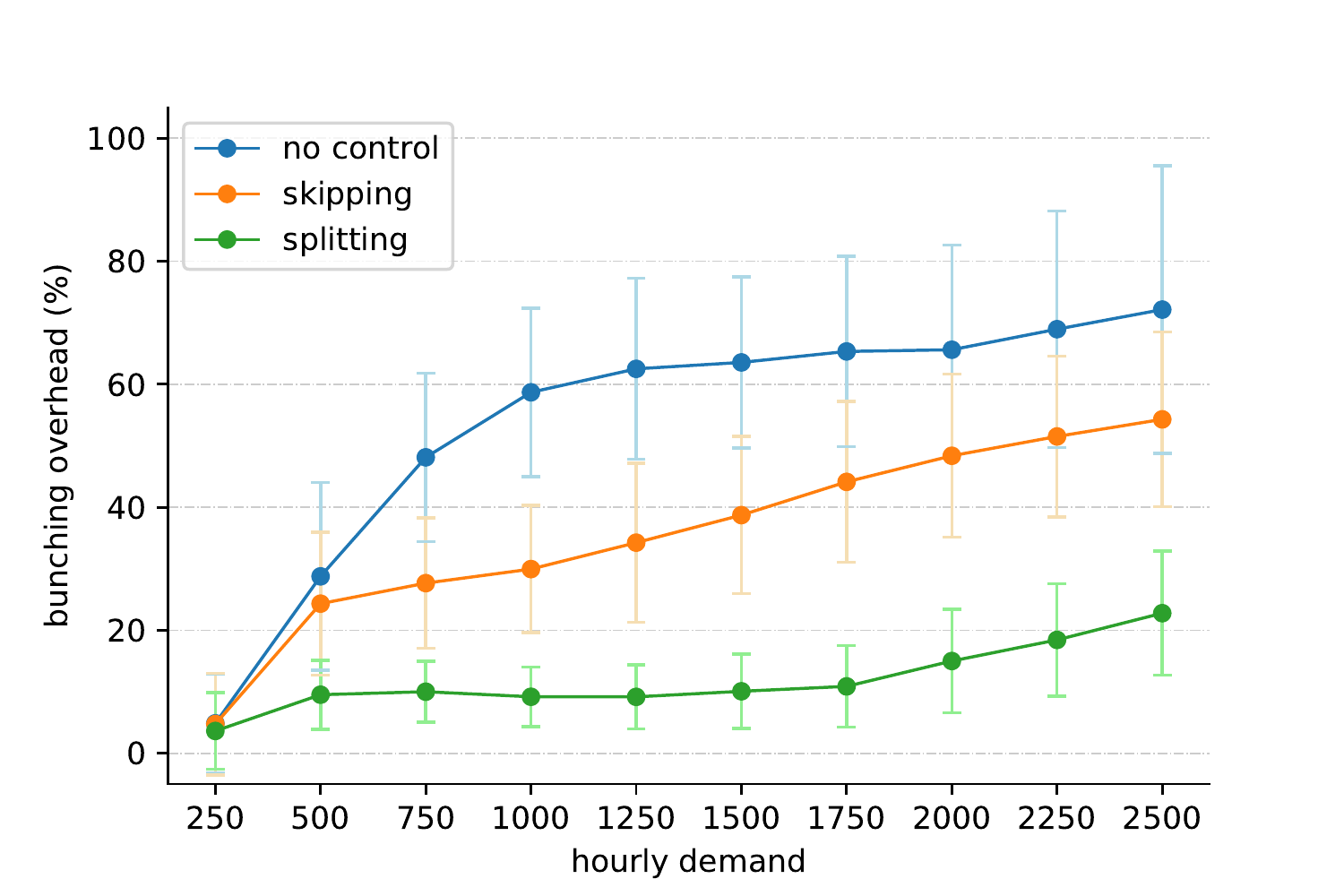}
        \caption{$\thresh=1.5$}\label{fig:overhead-thresh-15}
        \end{subfigure}
        \begin{subfigure}[b]{0.45\textwidth}\centering
        \includegraphics[width=\textwidth]{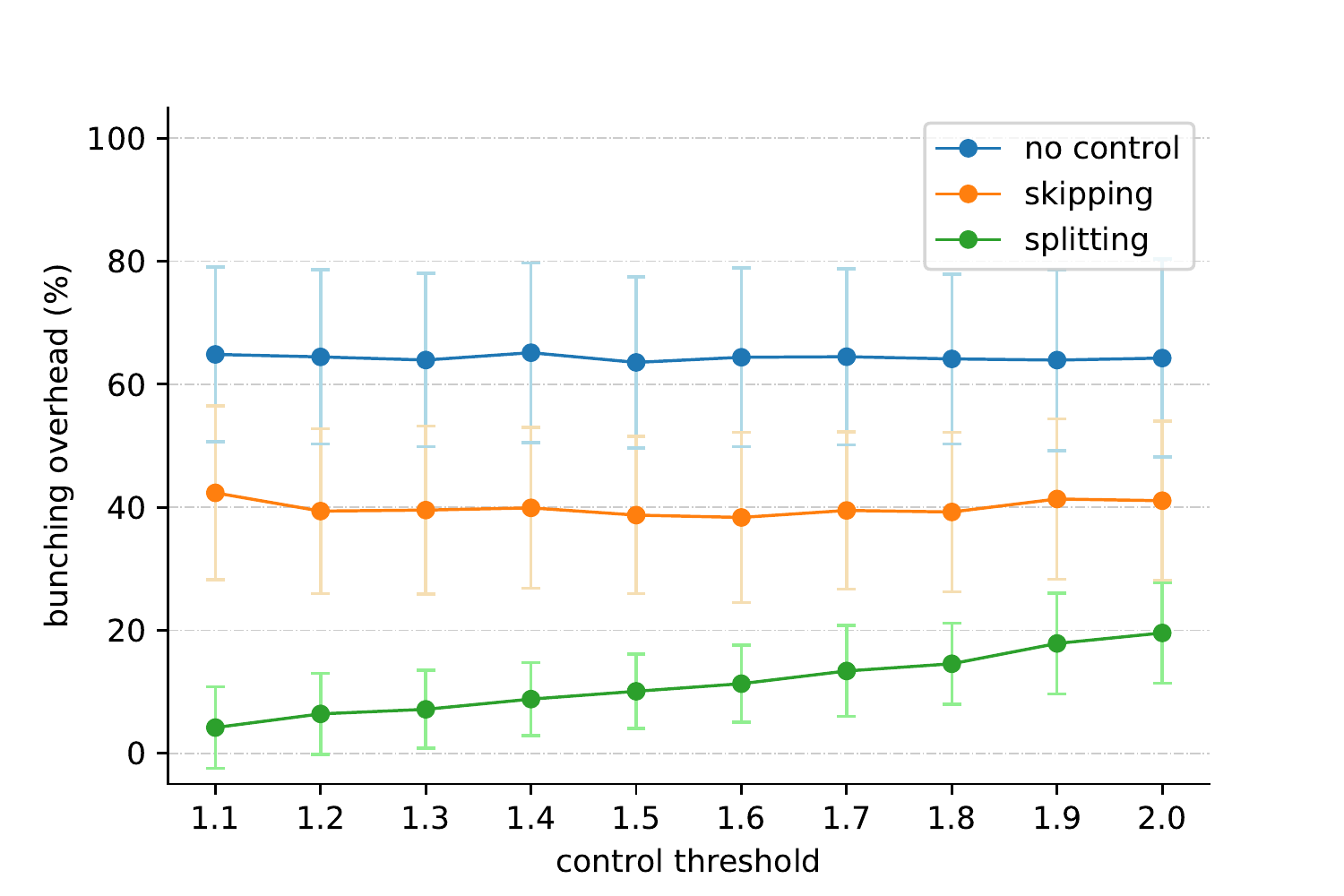}
        \caption{$\demand=1500$}\label{fig:overhead-demand-1500}
        \end{subfigure}
\caption{Bus bunching overhead $\overhead$ under various values of the control threshold $\thresh$ and demand $\demand$. Error bars show standard deviation.}\label{fig:overhead}
\end{figure}

Figure~\ref{fig:overhead} shows the bus bunching overhead $\overhead$, which is a measure of the inefficiency of the bus service due to headway variability. Figure~\ref{fig:overhead}(a) varies the demand while keeping the control threshold at $\thresh=1.5$, while Figure~\ref{fig:overhead}(b) varies the control threshold while keeping the demand at $\demand=1500$. The bus-splitting policy distinctly outperforms stop-skipping in all cases, often yielding more than double the reduction in the average travel cost as the latter (compared to no control). An alternative way of viewing this is that bus-splitting eliminates up to $50-75\%$ of the bunching overhead of stop-skipping. Figure~\ref{fig:overhead}(a) shows that bus-splitting maintains a stable $\overhead$ in the low-to-moderate demand range, whereas $\overhead$ is consistently increasing under the other two policies in this range. This clearly demonstrates that bus-splitting staves off most of the effects of bus bunching until the high demand range, where its $\overhead$ starts to increase. Even in the latter range, the $\overhead$ of bus-splitting remains well below half of that of stop-skipping. This suggests that replacing stop-skipping with bus-splitting is most beneficial for moderate-to-busy bus lines. Figure~\ref{fig:overhead}(b) shows that the choice of control threshold does not affect the performance of the stop-skipping policy. This is because a lower value of $\eta$ is more effective at reducing bus bunching, but the benefit is offset by the additional waiting and walking time imposed due to more stops being skipped. On the other hand, the bus-splitting policy is moderately sensitive to the choice of control threshold; a more proactive bus-splitting policy results in better performance. However, even the most conservative bus-splitting policy considered here decidedly outperforms stop-skipping.

\begin{figure}[!h]\centering
        \begin{subfigure}[b]{0.45\textwidth}\centering
        \includegraphics[width=\textwidth]{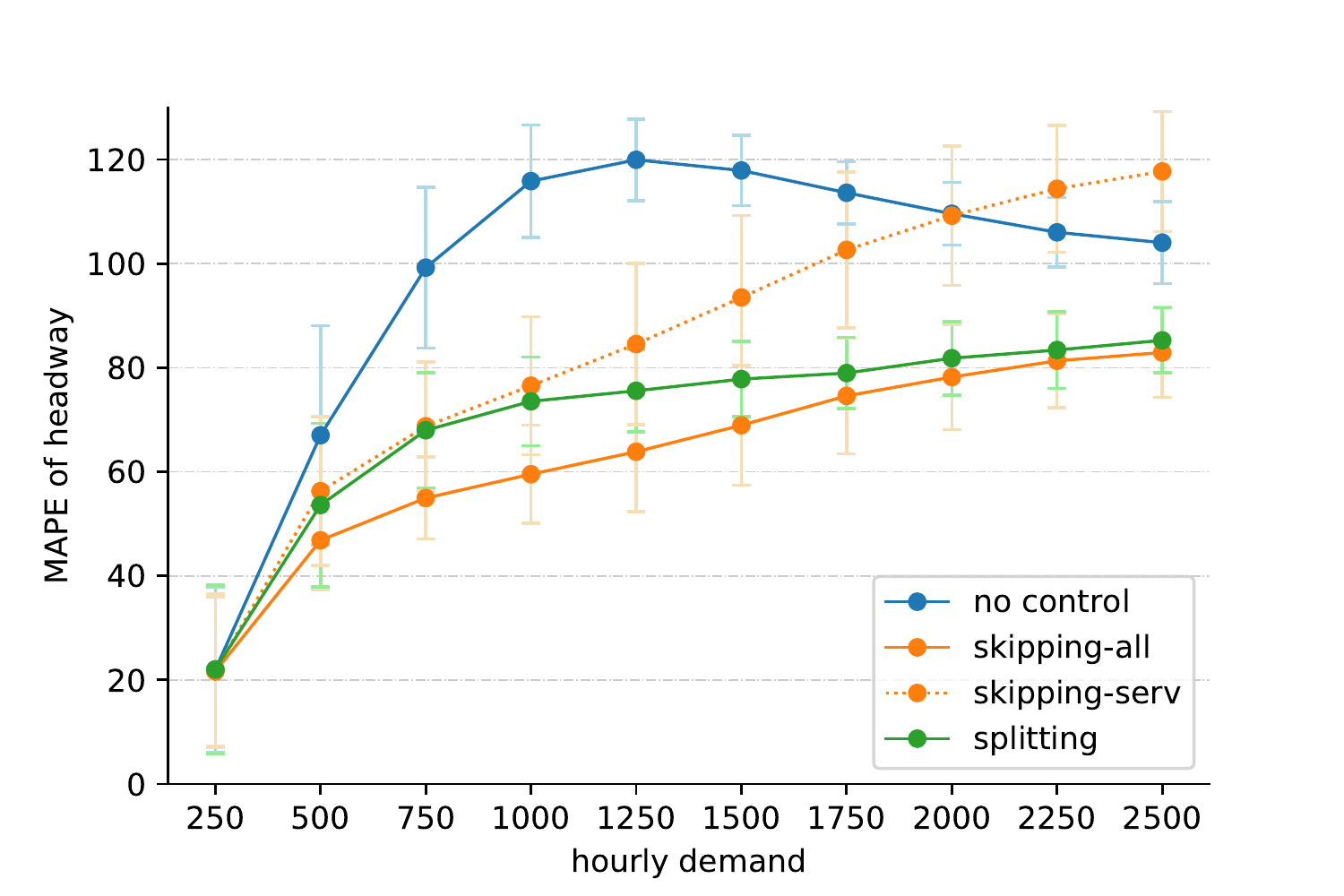}
        \caption{$\thresh=1.5$}\label{fig:mape-thresh-15}
        \end{subfigure}
        \begin{subfigure}[b]{0.45\textwidth}\centering
        \includegraphics[width=\textwidth]{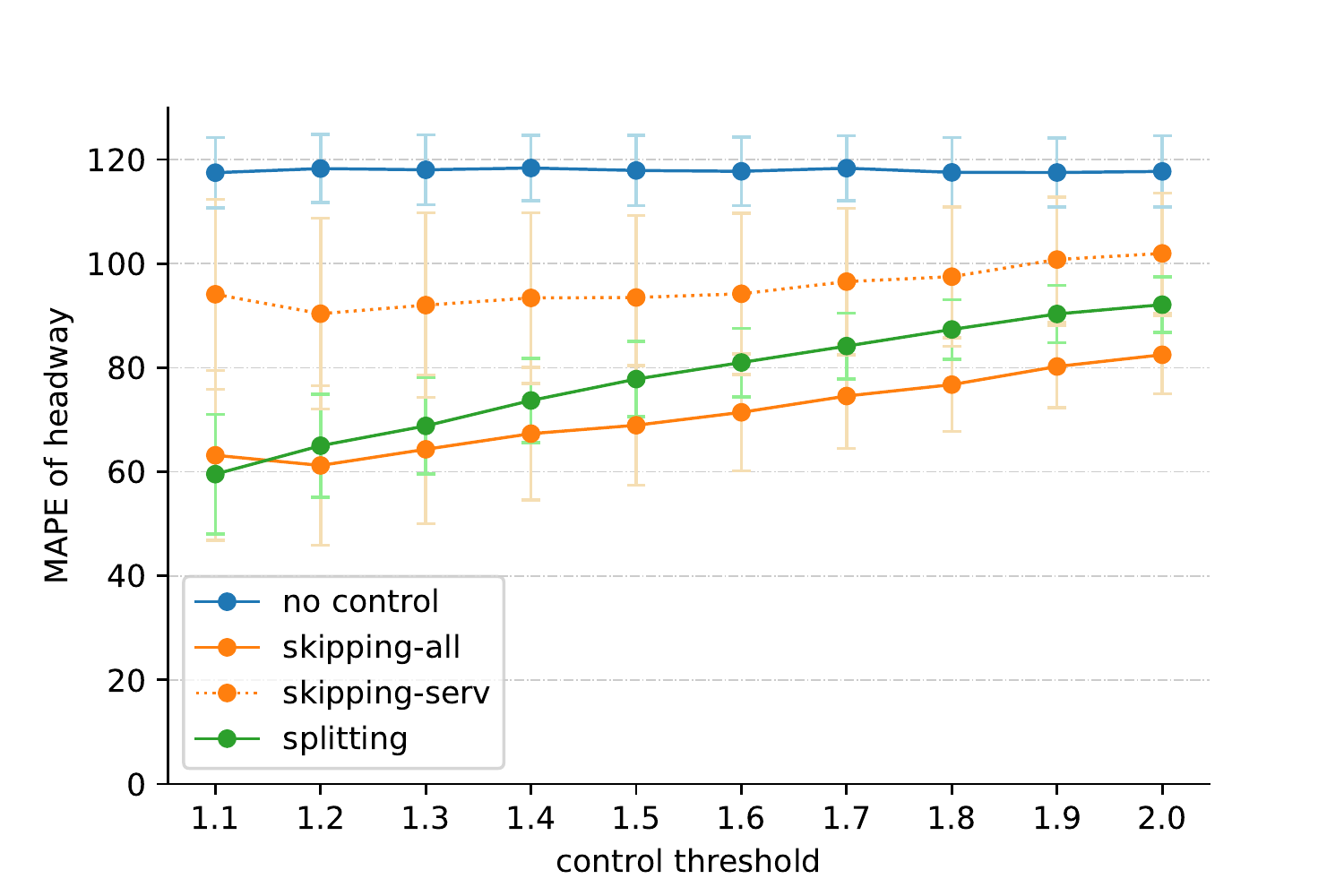}
        \caption{$\demand=1500$}\label{fig:mape-demand-1500}
        \end{subfigure}
\caption{Variability of the headway within experiment iterations under various values of the control threshold $\thresh$ and demand $\demand$, quantified by the mean absolute percentage error (MAPE) from the target headway $H$. Error bars show standard deviation.}\label{fig:mape}
\end{figure}

Figure~\ref{fig:mape} shows the variability of the headway within individual experiment iterations, quantified by the MAPE of the departing headway at each stop from the target headway $H$. The MAPE generally increases with both the demand and the control threshold. The former implies that the headways become more variable as demand increases, despite the fleet size being adjusted to account for the increase. However, this is also partially because the MAPE is a normalized metric with $H$ as the denominator, so the larger values of $H$ for smaller demands lead to lower values of the MAPE. The increase in the MAPE with the control threshold implies that a more proactive policy is more effective at reducing headway variability and therefore mitigating bus bunching. This is in agreement with our observations from Figure~\ref{fig:overhead}(b). The headway variability of the stop-skipping policy can be viewed from two perspectives. One approach is to consider the headways between consecutive buses regardless of whether they serve a given stop or not (solid orange line). Alternatively, one may account for the stops skipped by considering the headways between consecutive buses that do serve a given stop (dashed orange line). The MAPE of the former is consistently slightly lower than that under bus-splitting. This is due to the extra time incurred by one of the two units at the recoupling stop $s'_+$ while waiting for the other, which, compared to stop-skipping, delays the bus whose headway is long and needs to be reduced. We hypothesize that alternate flavors of bus-splitting, which do not require immediate recoupling, could mitigate this difference (albeit at the cost of imposing walking time on some passengers). The dashed orange line, on the other hand, represents the perspective of the passengers, who are indeed affected by the skipped stops. The MAPE of this headway is always higher than that of the bus-splitting policy, which helps explain why bus-splitting outperforms stop-skipping in terms of both the average travel cost in general and the average waiting time in particular.

\newpage

\begin{figure}[!h]\centering
      \begin{subfigure}[b]{0.45\textwidth}\centering
      \includegraphics[width=\textwidth]{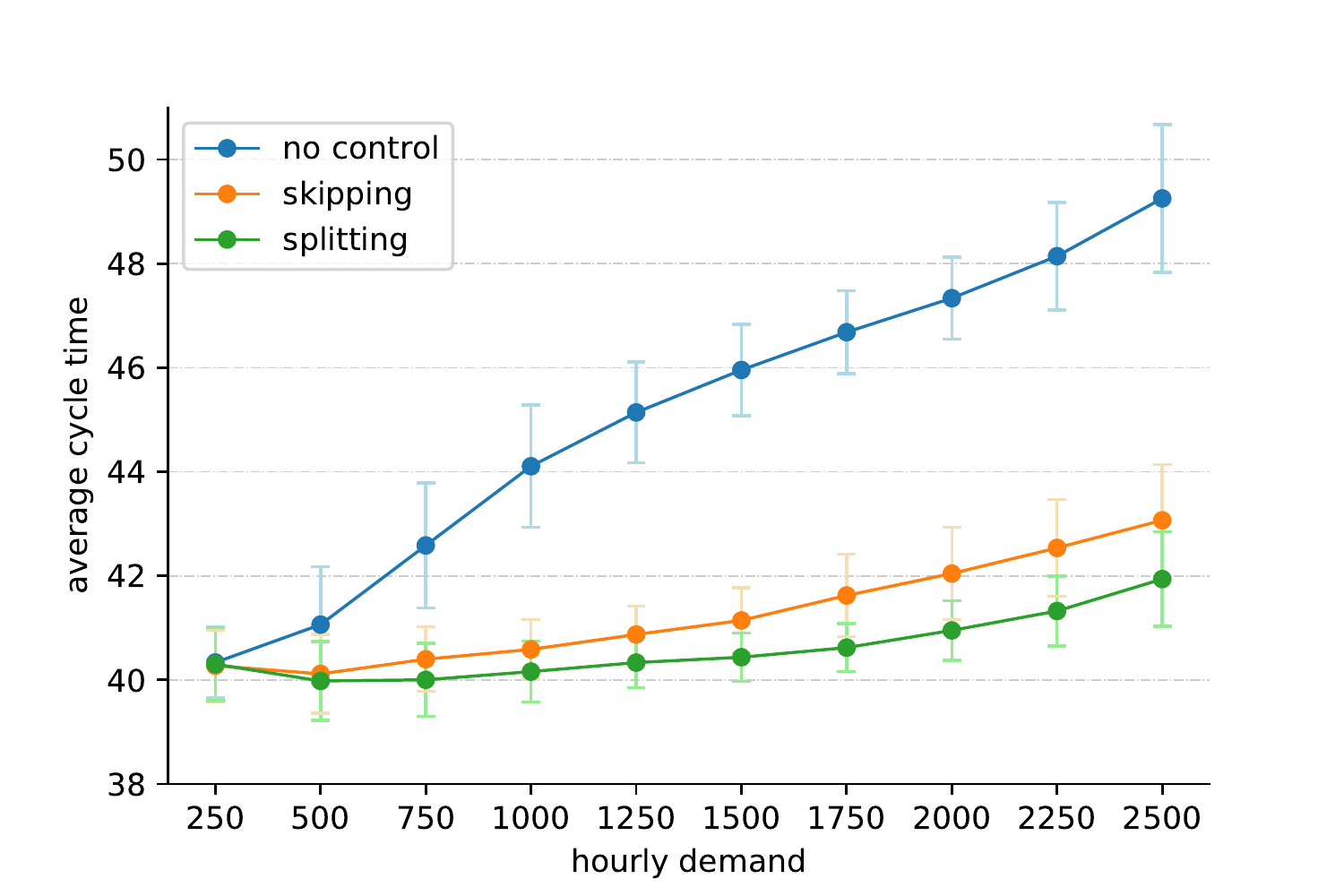}
      \caption{$\thresh=1.5$}\label{fig:cyc-thresh-15}
      \end{subfigure}
      \begin{subfigure}[b]{0.45\textwidth}\centering
      \includegraphics[width=\textwidth]{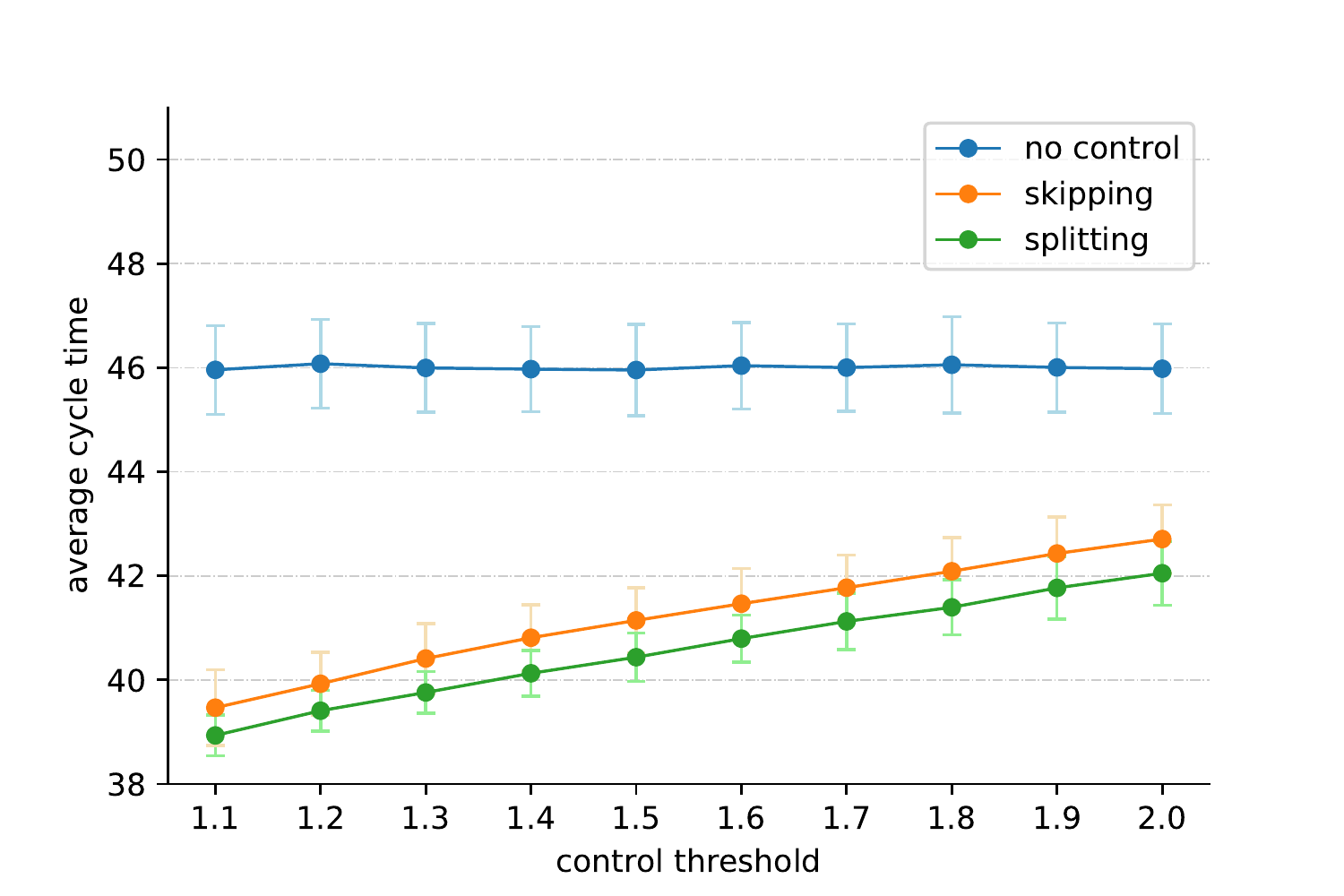}
      \caption{$\demand=1500$}\label{fig:cyc-demand-1500}
      \end{subfigure}
\caption{Average cycle length under various values of the control threshold $\thresh$ and demand $\demand$. The target cycle length is $\tau=40.5$ min. Error bars show standard deviation.}\label{fig:cycle}
\end{figure}

\begin{figure}[!h]\centering
      \begin{subfigure}[b]{0.45\textwidth}\centering
      \includegraphics[width=\textwidth]{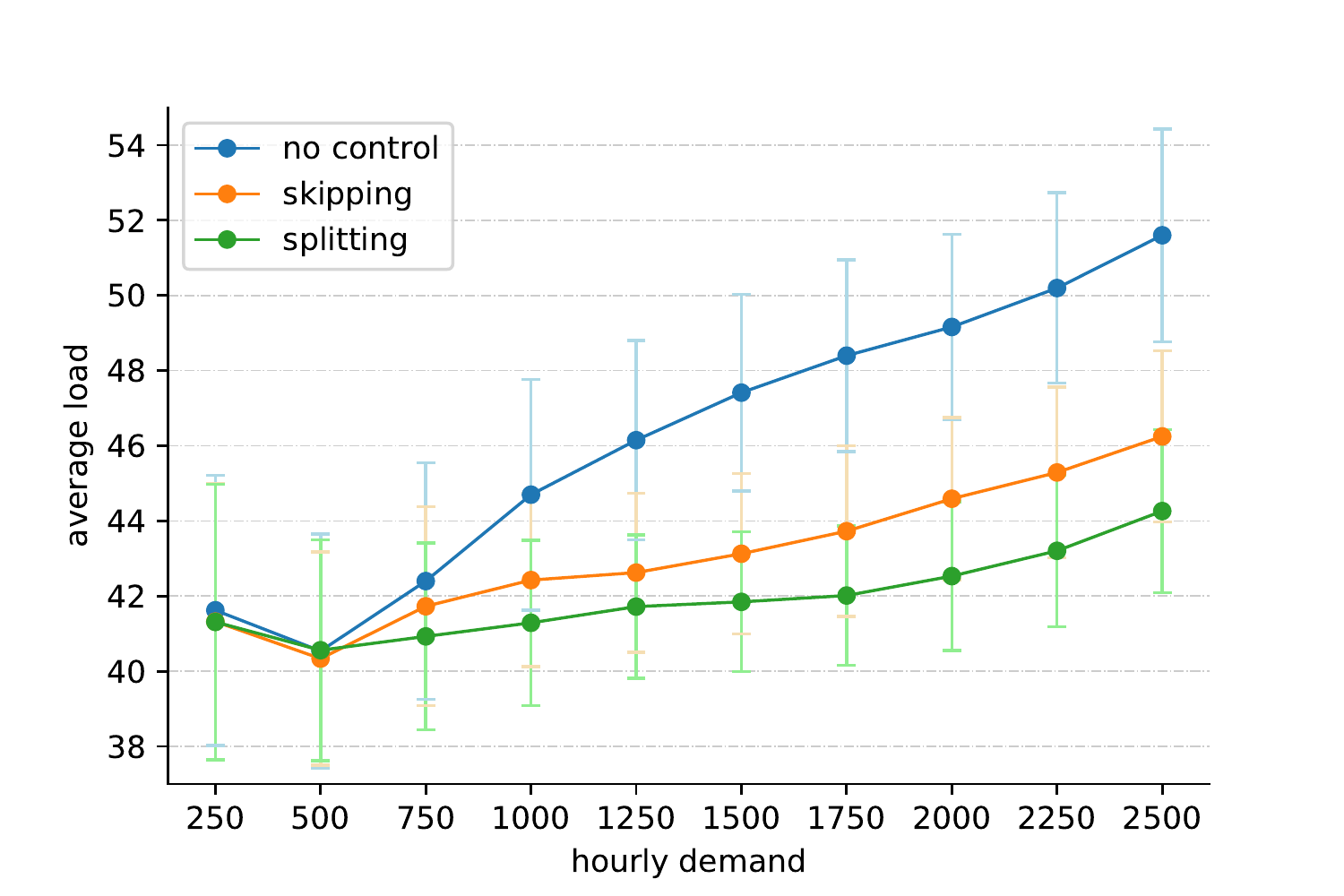}
      \caption{$\thresh=1.5$}\label{fig:load-thresh-15}
      \end{subfigure}
      \begin{subfigure}[b]{0.45\textwidth}\centering
      \includegraphics[width=\textwidth]{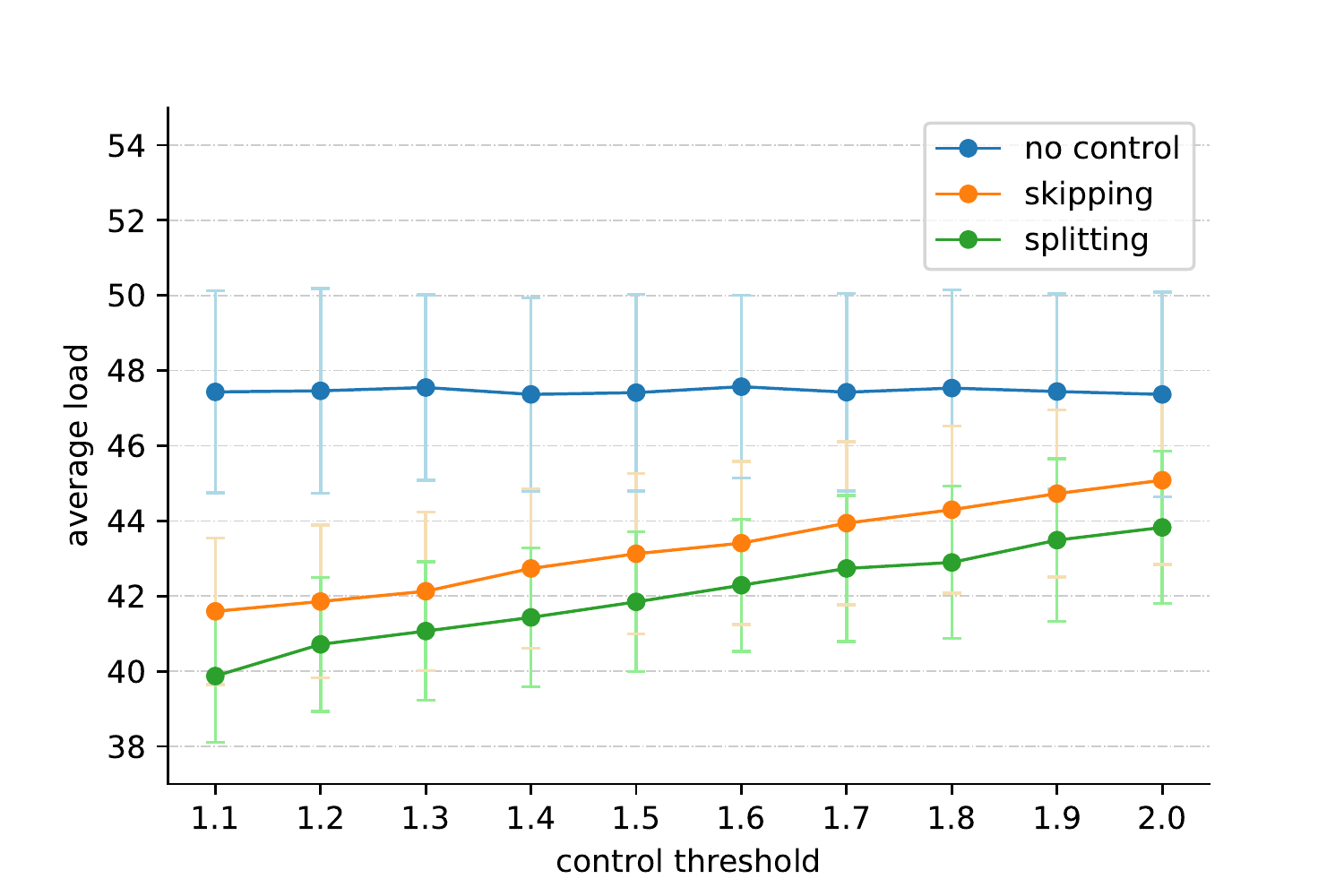}
      \caption{$\demand=1500$}\label{fig:load-demand-1500}
      \end{subfigure}
\caption{Average bus load under various values of the control threshold $\thresh$ and demand $\demand$. The target average load is $L=42$. Error bars show standard deviation.}\label{fig:load}
\end{figure}

Figure~\ref{fig:cycle} and Figure~\ref{fig:load} respectively show how the average cycle length and average bus load vary with the parameter values. These figures are discussed together since the trends they show are extremely similar. The cycle time and load under the bus-splitting policy are consistently slightly lower than under stop-skipping, which is in agreement with the average in-vehicle times under the two policies as seen in Figure~\ref{fig:costs}(b). Both policies have a significantly lower average cycle time and load than without control. The average cycle time and load increase with the demand. This reinforces our earlier observation that despite the increase in the fleet size to account for the increasing demand, the added variability of the headways is reducing the effective capacity of the system. This increase is much steeper without control, indicating that adopting either policy can allow a larger demand to be satisfied than without control. Figure~\ref{fig:cycle}(b) and Figure~\ref{fig:load}(b) show that the average cycle time and load also increase slowly with the control threshold. This is consistent with our observation from Figure~\ref{fig:mape}(b) that more proactive control policies are more effective.

\newpage

\begin{figure}[!h]\centering
      \begin{subfigure}[b]{0.45\textwidth}\centering
      \includegraphics[width=\textwidth]{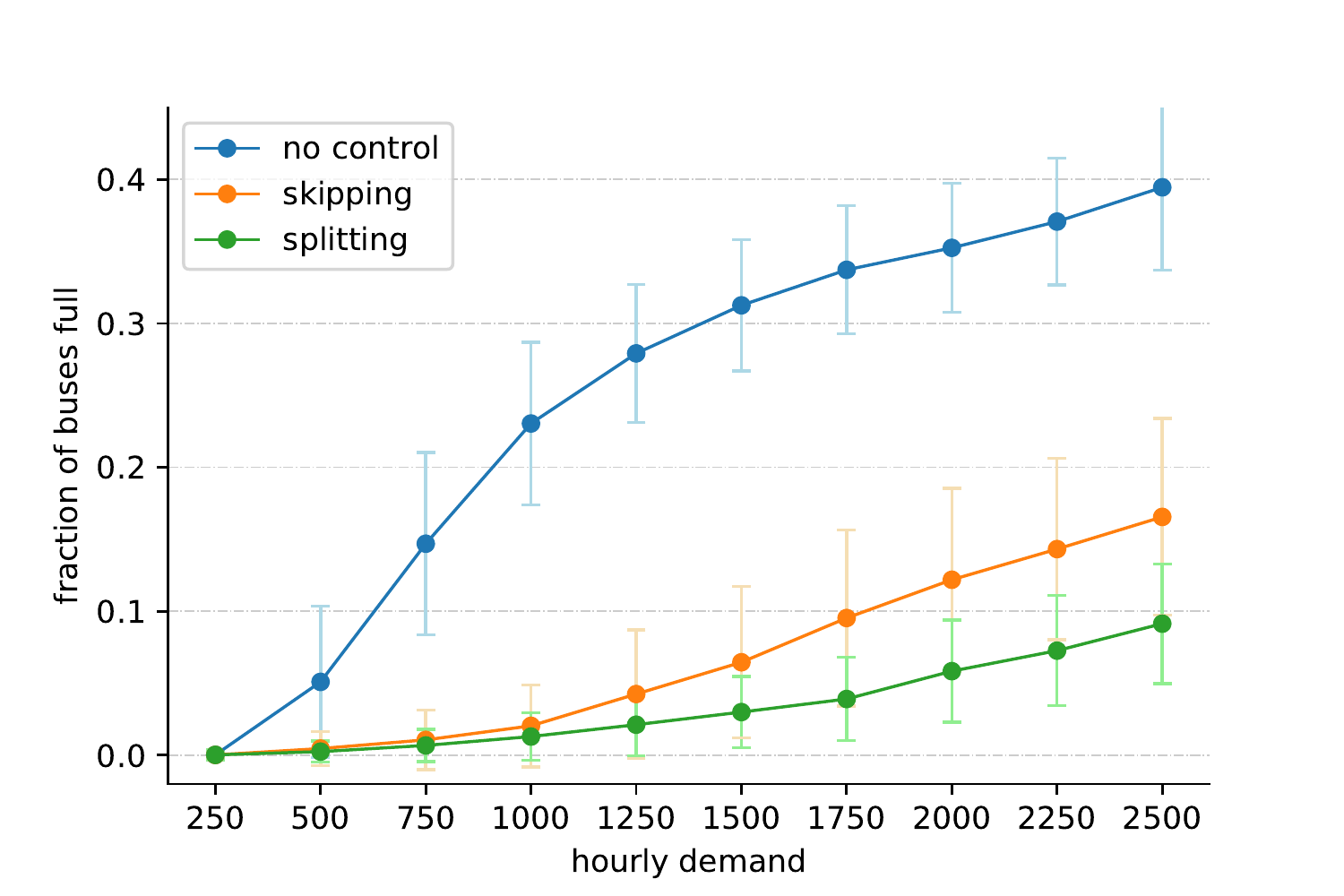}
      \caption{$\thresh=1.5$}\label{fig:fracfull-thresh-15}
      \end{subfigure}
      \begin{subfigure}[b]{0.45\textwidth}\centering
      \includegraphics[width=\textwidth]{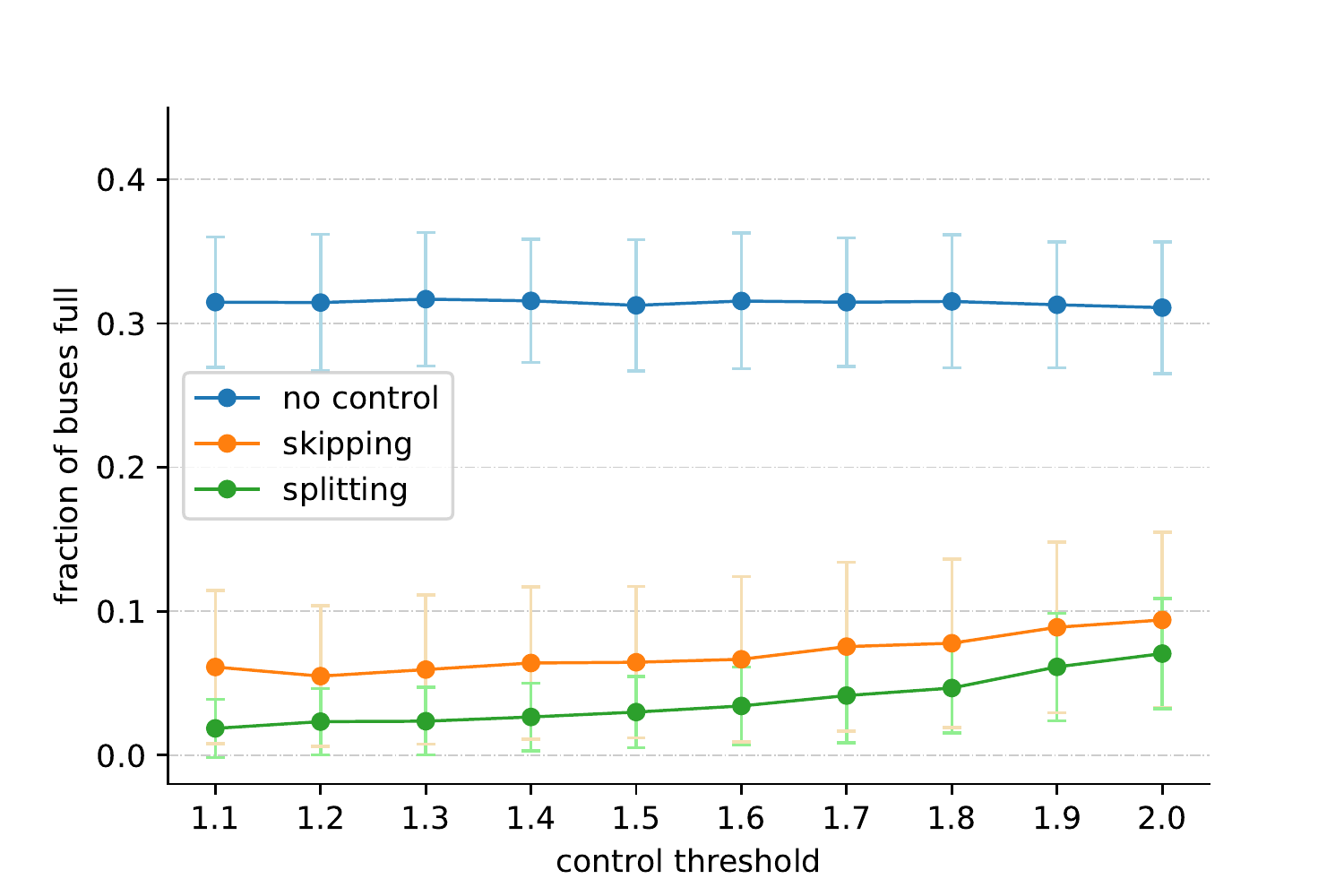}
      \caption{$\demand=1500$}\label{fig:fracfull-demand-1500}
      \end{subfigure}
\caption{Fraction of stops at which buses arrive full under various values of the control threshold $\thresh$ and demand $\demand$. Error bars show standard deviation.}\label{fig:fracfull}
\end{figure}

Figure~\ref{fig:fracfull} shows the fraction of stops at which buses arrive full. Both policies are able to maintain this fraction at a much lower value than under no control. This suggests that adopting either policy can allow a system to remain stable in cases where it would become oversaturated without control. The fraction rises slowly with both the demand and the control threshold. This mirrors our observations from Figures~\ref{fig:load}(a) and (b) regarding the decrease in effective capacity and the benefits of policy proactiveness respectively. Bus-splitting outperforms stop-skipping by a larger amount in terms of this metric in the high demand range. This is consistent with our observation from Figure~\ref{fig:overhead}(a) that the benefits of bus-splitting compared to stop-skipping are higher for busy bus lines.

The sensitivity analysis for the demand $\demand$ and the control threshold $\thresh$ shown above demonstrates that the proposed bus-splitting policy consistently offers significant travel cost savings to passengers by more effectively mitigating the overhead of bus bunching compared to the benchmark stop-skipping policy under a wide range of settings. In addition, we found that changing other parameter values, including the number of stops $S$, the fleet size factor $\fleetmult$, 
and the starting time and duration of the evaluation period, within reasonable limits, does not affect this main takeaway. 


\section{Conclusion \& Future Work}\label{sec:conclusion}

Autonomous modular vehicle (AMV) technology promises exciting capabilities for a wide range of transportation applications. With these technologies on the cusp of market availability, there is great interest in exploring the potential benefits they can bring to both fixed-line and flexible-route public transport. This paper is the first to study the use of autonomous modular buses (AMBs) to mitigate bus bunching. We present a simple ``bus-splitting" control strategy that directs a modular bus to decouple into two individual autonomous units when it experiences an unusually long headway. Each unit serves one stop while (partially or fully) skipping the other, so that the resources are allocated in parallel. This decreases the service time required, allowing the long headway to be shortened. We present an illustrative example to explain the concept of bus-splitting by comparing it with a traditional stop-skipping control strategy as well as a non-intervention (no control) strategy. We derive the dynamics of each strategy for a broadly applicable setting, and then use simulation to apply them to a specific setting consisting of a cyclical bus route with relatively homogeneous stops and demand. We evaluate the strategies in terms of the average travel cost faced by passengers as well as several other metrics for a wide range of parameter settings. We find that the proposed bus-splitting strategy reduces the overhead of bus bunching by (often more than) twice as much as the benchmark stop-skipping strategy, thereby offering significantly higher cost savings to passengers. This cost reduction is seen in each of the three travel time components: waiting time, in-vehicle time, and walking time (which it eliminates completely), and is highest for the levels of demand associated with busy bus lines. The proposed strategy also reduces headway variability to a comparable degree to the stop-skipping strategy. These results suggest that it is a superior alternative to stop-skipping for mitigating bus bunching. 
Furthermore, we analyze different thresholds for applying the proposed strategy, and show that it is most effective when applied proactively, i.e. with the control action being triggered even when the headway deviates from the target headway by a small amount.

Since the primary purpose of this work is to serve as a proof-of-concept for using AMBs to improve service reliability, we have limited our attention to a very basic flavor of modular bus-splitting. Our strategy, which is non-predictive, distributed, and myopic, can be improved and made more sophisticated in many ways. One example is adding a predictive element that anticipates the effect of the control action on future dynamics. Another is considering more information such as the load and location of upstream and downstream buses while making a control decision. Another is using machine learning and statistical techniques to determine the control action. Yet another is to analytically derive the optimal strategy. We can also examine the effect of relaxing several of our modeling assumptions. The foremost would be removing the requirement for modular units to recouple after two stops. We anticipate that this will improve the headway reliability at the cost of increasing the passenger travel cost. It is also important to consider how our methodology would need to be modified for asymmetric systems with certain very busy stops (i.e. transit hubs). 
Finally, we are keen to explore combining bus-splitting with other bus control strategies such as holding and bus insertion. We believe that this is a promising research area with rich insights to be gained into the operations and benefits of AMBs.


\section*{Acknowledgements}
    This work was supported by the NYUAD Center for Interacting Urban Networks (CITIES), funded by Tamkeen under the NYUAD Research Institute Award CG001. 








\bibliographystyle{Bibstyles/elsarticle-num}

\bibliography{refs}

\begin{thebibliography}{10}
\expandafter\ifx\csname url\endcsname\relax
  \def\url#1{\texttt{#1}}\fi
\expandafter\ifx\csname urlprefix\endcsname\relax\def\urlprefix{URL }\fi
\expandafter\ifx\csname href\endcsname\relax
  \def\href#1#2{#2} \def\path#1{#1}\fi

\bibitem{loder2017empirics}
A.~Loder, L.~Amb{\"u}hl, M.~Menendez, K.~W. Axhausen, Empirics of multi-modal
  traffic networks--using the 3d macroscopic fundamental diagram,
  Transportation Research Part C: Emerging Technologies 82 (2017) 88--101.

\bibitem{daganzo2009}
C.~F. Daganzo, A headway-based approach to eliminate bus bunching: Systematic
  analysis and comparisons, Transportation Research Part B: Methodological
  43~(10) (2009) 913--921.

\bibitem{newell1964}
G.~Newell, R.~Potts, Maintaining a bus schedule, in: Proceedings of the 2nd
  Australian Road Research Board, Vol.~43, 1964, pp. 388--393.

\bibitem{nesheli2015}
M.~M. Nesheli, A.~A. Ceder, Improved reliability of public transportation using
  real-time transfer synchronization, Transportation Research Part C: Emerging
  Technologies 60 (2015) 525--539.

\bibitem{daganzo2011}
C.~F. Daganzo, J.~Pilachowski, Reducing bunching with bus-to-bus cooperation,
  Transportation Research Part B: Methodological 45~(1) (2011) 267--277.

\bibitem{delgado2012}
F.~Delgado, J.~C. Munoz, R.~Giesen, How much can holding and/or limiting
  boarding improve transit performance?, Transportation Research Part B:
  Methodological 46~(9) (2012) 1202--1217.

\bibitem{berrebi2018}
S.~J. Berrebi, E.~Hans, N.~Chiabaut, J.~A. Laval, L.~Leclercq, K.~E. Watkins,
  Comparing bus holding methods with and without real-time predictions,
  Transportation Research Part C: Emerging Technologies 87 (2018) 197--211.

\bibitem{liu2013bus}
Z.~Liu, Y.~Yan, X.~Qu, Y.~Zhang, Bus stop-skipping scheme with random travel
  time, Transportation Research Part C: Emerging Technologies 35 (2013) 46--56.

\bibitem{niu2011determination}
H.~Niu, Determination of the skip-stop scheduling for a congested transit line
  by bilevel genetic algorithm, International journal of computational
  intelligence systems 4~(6) (2011) 1158--1167.

\bibitem{sun2005real}
A.~Sun, M.~Hickman, The real--time stop--skipping problem, Journal of
  Intelligent Transportation Systems 9~(2) (2005) 91--109.

\bibitem{petit2018dynamic}
A.~Petit, Y.~Ouyang, C.~Lei, Dynamic bus substitution strategy for bunching
  intervention, Transportation Research Part B: Methodological 115 (2018)
  1--16.

\bibitem{morales2019stochastic}
D.~Morales, J.~C. Mu{\~n}oz, P.~Gazmuri, A stochastic model for bus injection
  in a public transport service, Transportation Research Procedia 38 (2019)
  688--708.

\bibitem{menendez2021chapter}
M.~Menendez, Adaptive bus control, in: R.~Vickerman (Ed.), International
  Encyclopedia of Transportation, Vol.~4, Elsevier, 2021, pp. 315--324.

\bibitem{NextFTwebpage}
NextFutureTransportationInc., Next future transportation,
  http://www.next-future-mobility.com (2018, Accessed: 2020-02-14).

\bibitem{gecchelin2019modular}
T.~Gecchelin, J.~Webb, Modular dynamic ride-sharing transport systems, Economic
  Analysis and Policy 61 (2019) 111--117.

\bibitem{chen2019operational}
Z.~Chen, X.~Li, X.~Zhou, Operational design for shuttle systems with modular
  vehicles under oversaturated traffic: Discrete modeling method,
  Transportation Research Part B: Methodological 122 (2019) 1--19.

\bibitem{chen2020operational}
Z.~Chen, X.~Li, X.~Zhou, Operational design for shuttle systems with modular
  vehicles under oversaturated traffic: Continuous modeling method,
  Transportation Research Part B: Methodological 132 (2020) 76--100.

\bibitem{dai2020joint}
Z.~Dai, X.~C. Liu, X.~Chen, X.~Ma, Joint optimization of scheduling and
  capacity for mixed traffic with autonomous and human-driven buses: A dynamic
  programming approach, Transportation Research Part C: Emerging Technologies
  114 (2020) 598--619.

\bibitem{dakic2021design}
I.~Dakic, K.~Yang, M.~Menendez, J.~Y. Chow, On the design of an optimal
  flexible bus dispatching system with modular bus units: Using the
  three-dimensional macroscopic fundamental diagram, Transportation Research
  Part B: Methodological 148 (2021) 38--59.

\bibitem{chen2021designing}
Z.~Chen, X.~Li, Designing corridor systems with modular autonomous vehicles
  enabling station-wise docking: Discrete modeling method, Transportation
  Research Part E: Logistics and Transportation Review 152 (2021) 102388.

\bibitem{shi2020variable}
X.~Shi, Z.~Chen, M.~Pei, X.~Li, Variable-capacity operations with modular
  transits for shared-use corridors, Transportation Research Record 2674~(9)
  (2020) 230--244.

\bibitem{shi2021operations}
X.~Shi, X.~Li, Operations design of modular vehicles on an oversaturated
  corridor with first-in, first-out passenger queueing, Transportation Science.

\bibitem{wu2021modular}
J.~Wu, B.~Kulcs{\'a}r, X.~Qu, et~al., A modular, adaptive, and autonomous
  transit system (maats): A in-motion transfer strategy and performance
  evaluation in urban grid transit networks, Transportation Research Part A:
  Policy and Practice 151 (2021) 81--98.

\bibitem{caros2021day}
N.~S. Caros, J.~Y. Chow, Day-to-day market evaluation of modular autonomous
  vehicle fleet operations with en-route transfers, Transportmetrica B:
  Transport Dynamics 9~(1) (2021) 109--133.

\bibitem{gong2021transfer}
M.~Gong, Y.~Hu, Z.~Chen, X.~Li, Transfer-based customized modular bus system
  design with passenger-route assignment optimization, Transportation Research
  Part E: Logistics and Transportation Review 153 (2021) 102422.

\bibitem{vukan1973}
V.~R. Vuchic, Skip-stop operation as a method for transit speed increase,
  Traffic Quarterly (1956) 307--327.

\bibitem{tirachini2014economics}
A.~Tirachini, The economics and engineering of bus stops: Spacing, design and
  congestion, Transportation research part A: policy and practice 59 (2014)
  37--57.

\bibitem{kim2003performance}
H.~J. KIM, Performance of bus lanes in seoul: Some impacts and suggestions,
  IATSS Research 27~(2) (2003) 36--45.

\bibitem{chandra2013speed}
S.~Chandra, A.~K. Bharti, Speed distribution curves for pedestrians during
  walking and crossing, Procedia-Social and Behavioral Sciences 104 (2013)
  660--667.

\bibitem{kfh2013transit}
K.~Group, et~al., Transit capacity and quality of service manual.

\bibitem{wardman2004public}
M.~Wardman, Public transport values of time, Transport policy 11~(4) (2004)
  363--377.

\end{thebibliography}

\end{document}